\documentclass[apjl]{emulateapj}
\usepackage{comment}
\usepackage{ifthen}

\newcommand{\ctbd}[1]{}

\newcommand{\lc}{light curve}
\newcommand{\lcs}{light curves}
\newcommand{\Lc}{Light curve}

\newcommand{\band}[1]{\ensuremath{#1}~band}

\newcommand{\masy}{\ensuremath{\rm mas\,yr^{-1}}}
\newcommand{\kms}{\ensuremath{\rm km\,s^{-1}}}
\newcommand{\ms}{\ensuremath{\rm m\,s^{-1}}}

\newcommand{\gcmc}{\ensuremath{\rm g\,cm^{-3}}}
\newcommand{\ergscmsq}{\ensuremath{\rm erg\,s^{-1}\,cm^{-2}}}

\newcommand{\vsini}{\ensuremath{v \sin{i}}}
\newcommand{\feh}{\ensuremath{\rm [Fe/H]}}

\newcommand{\vmac}{\ensuremath{v_{\rm mac}}}
\newcommand{\vmic}{\ensuremath{v_{\rm mic}}}

\newcommand{\rsun}{\ensuremath{R_\sun}}
\newcommand{\msun}{\ensuremath{M_\sun}}
\newcommand{\lsun}{\ensuremath{L_\sun}}

\newcommand{\rstar}{\ensuremath{R_\star}}
\newcommand{\mstar}{\ensuremath{M_\star}}
\newcommand{\lstar}{\ensuremath{L_\star}}

\newcommand{\teffstar}{\ensuremath{T_{\rm eff\star}}}
\newcommand{\rhostar}{\ensuremath{\rho_\star}}
\newcommand{\loggstar}{\ensuremath{\log{g_{\star}}}}

\newcommand{\rearth}{\ensuremath{R_\earth}}
\newcommand{\mearth}{\ensuremath{M_\earth}}

\newcommand{\rpl}{\ensuremath{R_{p}}}
\newcommand{\mpl}{\ensuremath{M_{p}}}

\newcommand{\rhopl}{\ensuremath{\rho_{p}}}

\newcommand{\arstar}{\ensuremath{a/\rstar}}
\newcommand{\zrstar}{\ensuremath{\zeta/\rstar}}

\newcommand{\rjup}{\ensuremath{R_{\rm J}}}
\newcommand{\mjup}{\ensuremath{M_{\rm J}}}

\newcommand{\refsec}[1]{\mbox{\S\ \ref{sec:#1}}}

\newcommand{\reffigl}[1]{Figure~\ref{fig:#1}}
\newcommand{\refsecl}[1]{\mbox{Section \ref{sec:#1}}}

\newcommand{\reftabl}[1]{Table~\ref{tab:#1}}

\newcommand{\flwof}{\mbox{FLWO 1.2\,m}}

\newcommand{\hatcurfieldA}{376}                                         
\newcommand{\hatcurfieldB}{377}                                         
\newcommand{\hatcurCCra}{\ensuremath{14^{\mathrm h}12^{\mathrm m}37.44{\mathrm s}}}                                  
\newcommand{\hatcurCCdec}{\ensuremath{+04{\arcdeg}03{\arcmin}36.0{\arcsec}}}                                 
\newcommand{\hatcurCCtwomass}{2MASS~14123753+0403359}                  
\newcommand{\hatcurCCgsc}{GSC~0320-01027}                              
\newcommand{\hatcurCCtassmv}{11.744}                                   
\newcommand{\hatcurCCtwomassJmag}{\ensuremath{10.080\pm0.022}}         
\newcommand{\hatcurCCtwomassHmag}{\ensuremath{9.685\pm0.023}}          
\newcommand{\hatcurCCtwomassKmag}{\ensuremath{9.581\pm0.023}}          
\newcommand{\hatcurCCesoJKmag}{\ensuremath{0.530\pm0.035}}             
\newcommand{\hatcurLCdip}{\ensuremath{4.9}}                            
\newcommand{\hatcurLCrprstar}{\ensuremath{0.0737\pm0.0012}}            
\newcommand{\hatcurLCbsq}{\ensuremath{0.092_{-0.053}^{+0.087}}}        
\newcommand{\hatcurLCimp}{\ensuremath{0.303_{-0.122}^{+0.112}}}        
\newcommand{\hatcurLCzeta}{\ensuremath{21.15\pm0.16}}                  
\newcommand{\hatcurLCdur}{\ensuremath{0.1023\pm0.0010}}                
\newcommand{\hatcurLCdurhr}{\ensuremath{2.455\pm0.025}}                
\newcommand{\hatcurLCq}{\ensuremath{0.0242\pm0.0002}}                  
\newcommand{\hatcurLCingdur}{\ensuremath{0.0077\pm0.0007}}             
\newcommand{\hatcurLCP}{\ensuremath{4.234516\pm0.000015}}              
\newcommand{\hatcurLCPprec}{\ensuremath{4.2345156}}                    
\newcommand{\hatcurLCPshort}{\ensuremath{4.2345}}                      
\newcommand{\hatcurLCT}{\ensuremath{2455304.65122\pm0.00035}}          
\newcommand{\hatcurLCTA}{\ensuremath{2454860.02709\pm0.00147}}         
\newcommand{\hatcurLCTB}{\ensuremath{2455342.76185\pm0.00041}}         
\newcommand{\hatcurLCiblendA}{\ensuremath{0.92\pm0.05}}                
\newcommand{\hatcurLCiblendB}{\ensuremath{0.84\pm0.10}}                
\newcommand{\hatcurSMEiteff}{\ensuremath{5079\pm88}}                   
\newcommand{\hatcurSMEiteffshort}{\ensuremath{5079}}                   
\newcommand{\hatcurSMEizfeh}{\ensuremath{-0.04\pm0.08}}                
\newcommand{\hatcurSMEizfehshort}{\ensuremath{-0.04}}                  
\newcommand{\hatcurSMEilogg}{\ensuremath{4.53\pm0.06}}                 
\newcommand{\hatcurSMEivsin}{\ensuremath{1.8\pm0.5}}                   
\newcommand{\hatcurSMEivmac}{\ensuremath{2.95}}                        
\newcommand{\hatcurSMEivmic}{\ensuremath{0.85}}                        
\newcommand{\hatcurSMEiiteff}{\ensuremath{5079\pm88}}                  
\newcommand{\hatcurSMEiizfeh}{\ensuremath{-0.04\pm0.08}}               
\newcommand{\hatcurSMEiizfehshort}{\ensuremath{-0.04}}                 
\newcommand{\hatcurSMEiilogg}{\ensuremath{4.53\pm0.06}}                
\newcommand{\hatcurSMEiivsin}{\ensuremath{1.8\pm0.5}}                  
\newcommand{\hatcurSMEiivmac}{\ensuremath{NULL}}                       
\newcommand{\hatcurSMEiivmic}{\ensuremath{NULL}}                       
\newcommand{\hatcurTRESteff}{\ensuremath{5125\pm125}}                  
\newcommand{\hatcurTRESlogg}{\ensuremath{4.5\pm0.5}}                   
\newcommand{\hatcurTRESvsini}{\ensuremath{1.0\pm1.0}}                    
\newcommand{\hatcurTRESgamma}{\ensuremath{+14.72\pm0.10}}              
\newcommand{\hatcurLBii}{\ensuremath{0.3862}}                          
\newcommand{\hatcurLBiii}{\ensuremath{0.2576}}                         
\newcommand{\hatcurISOm}{\ensuremath{0.82\pm0.03}}                     
\newcommand{\hatcurISOmlong}{\ensuremath{0.816\pm0.033}}               
\newcommand{\hatcurISOr}{\ensuremath{0.79_{-0.04}^{+0.10}}}            
\newcommand{\hatcurISOrlong}{\ensuremath{0.788_{-0.043}^{+0.098}}}     
\newcommand{\hatcurISOlogg}{\ensuremath{4.56\pm0.06}}                  
\newcommand{\hatcurISOlum}{\ensuremath{0.38_{-0.06}^{+0.16}}}          
\newcommand{\hatcurISOmv}{\ensuremath{6.03\pm0.24}}                    
\newcommand{\hatcurISOage}{\ensuremath{9.0_{-4.9}^{+3.0}}}             
\newcommand{\hatcurISOMK}{\ensuremath{3.98\pm0.19}}                    
\newcommand{\hatcurISOJK}{\ensuremath{0.55\pm0.02}}                    
\newcommand{\hatcurISOspec}{K1}                                        
\newcommand{\hatcurRVK}{\ensuremath{8.5\pm1.0}}                        
\newcommand{\hatcurRVk}{\ensuremath{0.099\pm0.060}}                    
\newcommand{\hatcurRVh}{\ensuremath{0.027\pm0.076}}                    
\newcommand{\hatcurRVgamma}{\ensuremath{-2.8\pm0.8}}                   
\newcommand{\hatcurRVjitter}{\ensuremath{1.6}}                         
\newcommand{\hatcurRVeccen}{\ensuremath{0.124\pm0.060}}                
\newcommand{\hatcurRVomega}{\ensuremath{54\pm165}}                     
\newcommand{\hatcurRVfitrms}{\ensuremath{2.4}}                         
\newcommand{\hatcurPPi}{\ensuremath{88.6_{-0.9}^{+0.5}}}               
\newcommand{\hatcurPPlogg}{\ensuremath{2.65_{-0.10}^{+0.08}}}          
\newcommand{\hatcurPPar}{\ensuremath{13.06\pm0.83}}                    
\newcommand{\hatcurPParel}{\ensuremath{0.0479\pm0.0006}}               
\newcommand{\hatcurPPrho}{\ensuremath{0.40\pm0.10}}                    
\newcommand{\hatcurPPmshort}{\ensuremath{0.06}}                        
\newcommand{\hatcurPPmlong}{\ensuremath{0.059\pm0.007}}                
\newcommand{\hatcurPPrshort}{\ensuremath{0.57}}                        
\newcommand{\hatcurPPrlong}{\ensuremath{0.565_{-0.032}^{+0.072}}}      
\newcommand{\hatcurPPmrcorr}{\ensuremath{0.07}}                        
\newcommand{\hatcurPPteff}{\ensuremath{1001_{-37}^{+66}}}              
\newcommand{\hatcurPPtheta}{\ensuremath{0.012\pm0.002}}                
\newcommand{\hatcurPPfluxperi}{\ensuremath{2.91_{-0.48}^{+7.54}}}      
\newcommand{\hatcurPPfluxperidim}{\ensuremath{8}}                      
\newcommand{\hatcurPPfluxap}{\ensuremath{1.81\pm0.32}}                 
\newcommand{\hatcurPPfluxapdim}{\ensuremath{8}}                        
\newcommand{\hatcurPPfluxavg}{\ensuremath{2.27_{-0.31}^{+1.08}}}       
\newcommand{\hatcurPPfluxavgdim}{\ensuremath{8}}                       
\newcommand{\hatcurXsecondary}{\ensuremath{2455307.037\pm0.162}}       
\newcommand{\hatcurXsecdur}{\ensuremath{0.1074\pm0.0162}}              
\newcommand{\hatcurXsecingdur}{\ensuremath{0.0082\pm0.0067}}           
\newcommand{\hatcurXdist}{\ensuremath{134_{-8}^{+18}}}                 
\newcommand{\hatcurCCpm}{\ensuremath{148.5\pm2.7}}               

\newcommand{\hatcur}{HAT-P-26}
\newcommand{\hatcurb}{HAT-P-26b}

\newcommand{\hatcurCCtassvi}{\ensuremath{0.96\pm0.11}}                  
\newcommand{\hatcurSMEversion}{i}                                       
\newcommand{\hatcurSMEteff}{\ifthenelse{\equal{\hatcurSMEversion}{i}}{\hatcurSMEiteff}{\hatcurSMEiiteff}}
\newcommand{\hatcurSMEzfeh}{\ifthenelse{\equal{\hatcurSMEversion}{i}}{\hatcurSMEizfeh}{\hatcurSMEiizfeh}}
\newcommand{\hatcurSMEzfehshort}{\ifthenelse{\equal{\hatcurSMEversion}{i}}{\hatcurSMEizfehshort}{\hatcurSMEiizfehshort}}
\newcommand{\hatcurSMElogg}{\ifthenelse{\equal{\hatcurSMEversion}{i}}{\hatcurSMEilogg}{\hatcurSMEiilogg}}
\newcommand{\hatcurSMEvsin}{\ifthenelse{\equal{\hatcurSMEversion}{i}}{\hatcurSMEivsin}{\hatcurSMEiivsin}}
\newcommand{\hatcurSMEvmac}{\ifthenelse{\equal{\hatcurSMEversion}{i}}{\hatcurSMEivmac}{\hatcurSMEiivmac}}
\newcommand{\hatcurSMEvmic}{\ifthenelse{\equal{\hatcurSMEversion}{i}}{\hatcurSMEivmic}{\hatcurSMEiivmic}}

\newcommand{\hatcurnoeccenLCrprstar}{\ensuremath{0.0738\pm0.0012}}     
\newcommand{\hatcurnoeccenLCbsq}{\ensuremath{0.110_{-0.059}^{+0.076}}} 
\newcommand{\hatcurnoeccenLCimp}{\ensuremath{0.332_{-0.123}^{+0.095}}} 
\newcommand{\hatcurnoeccenLCzeta}{\ensuremath{21.14\pm0.16}}           
\newcommand{\hatcurnoeccenLCdur}{\ensuremath{0.1025\pm0.0010}}         
\newcommand{\hatcurnoeccenLCingdur}{\ensuremath{0.0078\pm0.0007}}      
\newcommand{\hatcurnoeccenLCP}{\ensuremath{4.234515\pm0.000015}}       
\newcommand{\hatcurnoeccenLCT}{\ensuremath{2455304.65118\pm0.00036}}   
\newcommand{\hatcurnoeccenLBii}{\ensuremath{0.3862}}                   
\newcommand{\hatcurnoeccenLBiii}{\ensuremath{0.2576}}                  
\newcommand{\hatcurnoeccenRVK}{\ensuremath{8.3\pm1.0}}                 
\newcommand{\hatcurnoeccenRVk}{\ensuremath{0.000\pm0.000}}             
\newcommand{\hatcurnoeccenRVh}{\ensuremath{0.000\pm0.000}}             
\newcommand{\hatcurnoeccenRVjitter}{\ensuremath{2.4}}                  
\newcommand{\hatcurnoeccenRVeccen}{\ensuremath{0.000\pm0.000}}         
\newcommand{\hatcurnoeccenRVomega}{\ensuremath{0\pm0}}                 
\newcommand{\hatcurnoeccenRVfitrms}{\ensuremath{3.0}}                  
\newcommand{\hatcurnoeccenPPi}{\ensuremath{88.6\pm0.5}}                
\newcommand{\hatcurnoeccenPPlogg}{\ensuremath{2.67\pm0.07}}            
\newcommand{\hatcurnoeccenPPar}{\ensuremath{13.44_{-0.59}^{+0.44}}}    
\newcommand{\hatcurnoeccenPParel}{\ensuremath{0.0478\pm0.0006}}        
\newcommand{\hatcurnoeccenPPrho}{\ensuremath{0.42\pm0.08}}             
\newcommand{\hatcurnoeccenPPmlong}{\ensuremath{0.057\pm0.007}}         
\newcommand{\hatcurnoeccenPPrlong}{\ensuremath{0.549_{-0.023}^{+0.034}}} 
\newcommand{\hatcurnoeccenPPmrcorr}{\ensuremath{0.08}}                 
\newcommand{\hatcurnoeccenPPteff}{\ensuremath{981\pm29}}               
\newcommand{\hatcurnoeccenPPtheta}{\ensuremath{0.012\pm0.002}}         
\newcommand{\hatcurnoeccenPPfluxperi}{\ensuremath{2.10_{-0.20}^{+0.30}}}      
\newcommand{\hatcurnoeccenPPfluxap}{\ensuremath{2.10_{-0.20}^{+0.30}}}        
\newcommand{\hatcurnoeccenPPfluxavg}{\ensuremath{2.10_{-0.20}^{+0.30}}}       
\newcommand{\hatcurnoeccenXsecondary}{\ensuremath{2455306.768\pm0.000}} 
\newcommand{\hatcurnoeccenXsecdur}{\ensuremath{0.1025\pm0.0010}}       
\newcommand{\hatcurnoeccenXsecingdur}{\ensuremath{0.0078\pm0.0007}}    

\newcommand{\hatcurisoshort}{YY}
\newcommand{\hatcurisofull}{Yonsei-Yale (YY)}
\newcommand{\hatcurisocite}{yi:2001}
\newcommand{\hatcurlumind}{\arstar}
\newcommand{\hatcurjhkfilset}{ESO}

\newcommand{\hatcurRHKgyroage}{\ensuremath{7.8^{+1.4}_{-1.2}}}
\newcommand{\hatcurRHKlogage}{\ensuremath{9.80\pm0.15}}
\newcommand{\hatcurRHKage}{\ensuremath{6.4^{+2.7}_{-1.9}}}
\newcommand{\hatcurRHKro}{\ensuremath{2.2\pm0.2}}
\newcommand{\hatcurRHKprot}{\ensuremath{48\pm4}}
\newcommand{\hatcurRHKrhkmed}{\ensuremath{-4.992}}
\newcommand{\hatcurRHKsmed}{\ensuremath{0.182}}
\newcommand{\hatcurRHKbv}{\ensuremath{0.89}}
\newcommand{\hatcurRHKsrms}{\ensuremath{0.004}}
\newcommand{\hatcurRHKsbaselevel}{\ensuremath{0.168}}
\newcommand{\hatcurRHKprotvsini}{\ensuremath{22.3^{+17.2}_{-4.6}}}

\newboolean{emulateapj}
\setboolean{emulateapj}{true}

\newboolean{rvtablelong}
\setboolean{rvtablelong}{true}

\newboolean{astroph}
\setboolean{astroph}{true}

\shortauthors{Hartman et al.}
\shorttitle{\hatcur\lowercase{b}}
\ifthenelse{\boolean{emulateapj}}{
    \newcommand{\titledag}{$\dagger$}
}{
    \newcommand{\titledag}{\dagger}
}

\begin{document}

\title{\hatcur\lowercase{b}: A Low-Density Neptune-Mass
Planet Transiting a K Star\altaffilmark{\titledag}}

\author{
        J.~D.~Hartman\altaffilmark{1},
        G.~\'A.~Bakos\altaffilmark{1,2},
        D.~M.~Kipping\altaffilmark{1,3},
        G.~Torres\altaffilmark{1},
        G.~Kov\'acs\altaffilmark{4},
        R.~W.~Noyes\altaffilmark{1},
        D.~W.~Latham\altaffilmark{1},
        A.~W.~Howard\altaffilmark{5},
        D.~A.~Fischer\altaffilmark{6},
        J.~A.~Johnson\altaffilmark{7},
        G.~W.~Marcy\altaffilmark{5},
        H.~Isaacson\altaffilmark{5},
        S.~N.~Quinn\altaffilmark{1},
        L.~A.~Buchhave\altaffilmark{1,8},
        B.~B\'eky\altaffilmark{1},
        D.~D.~Sasselov\altaffilmark{1},
        R.~P.~Stefanik\altaffilmark{1},
        G.~A.~Esquerdo\altaffilmark{1},
        M.~Everett\altaffilmark{1},
        G.~Perumpilly\altaffilmark{1,9},
        J.~L\'az\'ar\altaffilmark{10},
        I.~Papp\altaffilmark{10},
        P.~S\'ari\altaffilmark{10}
}
\altaffiltext{1}{Harvard-Smithsonian Center for Astrophysics,
	Cambridge, MA; email: gbakos@cfa.harvard.edu}

\altaffiltext{2}{NSF Fellow}

\altaffiltext{3}{University College London, Dept. of Physics and
  Astronomy, Gower St., London, UK}

\altaffiltext{4}{Konkoly Observatory, Budapest, Hungary}

\altaffiltext{5}{Department of Astronomy, University of California,
	Berkeley, CA}

\altaffiltext{6}{Department of Astronomy, Yale University, New Haven, CT}

\altaffiltext{7}{Department of Astrophysics, California Institute of
  Technology, Pasadena, CA}

\altaffiltext{8}{Niels Bohr Institute, Copenhagen University, DK-2100,
  Copenhagen, Denmark}

\altaffiltext{9}{Department of Physics, University of South Dakota,
  Vermillion, SD}

\altaffiltext{10}{Hungarian Astronomical Association, Budapest, 
	Hungary}

\altaffiltext{$\dagger$}{
	Based in part on observations obtained at the W.~M.~Keck
	Observatory, which is operated by the University of California and
	the California Institute of Technology. Keck time has been
	granted by NASA (N018Hr and N167Hr).
}


\begin{abstract}

\setcounter{footnote}{10}

We report the discovery of \hatcurb{}, a transiting extrasolar planet
orbiting the moderately bright
V=\hatcurCCtassmv\ \hatcurISOspec\ dwarf star \hatcurCCgsc, with a
period $P=\hatcurLCP$\,d, transit epoch $T_c = \hatcurLCT$
(BJD\footnote{Barycentric Julian dates throughout the paper are
  calculated from Coordinated Universal Time (UTC)}), and transit
duration \hatcurLCdur\,d.  The host star has a mass of
\hatcurISOm\,\msun, radius of \hatcurISOr\,\rsun, effective
temperature \hatcurSMEteff\,K, and metallicity $\feh =
\hatcurSMEzfeh$.  The planetary companion has a mass of
\hatcurPPmlong\,\mjup, and radius of \hatcurPPrlong\,\rjup\ yielding a
mean density of \hatcurPPrho\,\gcmc. \hatcurb{} is the fourth
Neptune-mass transiting planet discovered to date. It has a mass that
is comparable to those of Neptune and Uranus, and slightly smaller
than those of the other transiting Super-Neptunes, but a radius that
is $\sim$65\% larger than those of Neptune and Uranus, and also larger
than those of the other transiting Super-Neptunes. \hatcurb{} is
consistent with theoretical models of an irradiated Neptune-mass
planet with a 10\,\mearth\ heavy element core that comprises $\ga
50\%$ of its mass with the remainder contained in a significant
hydrogen-helium envelope, though the exact composition is uncertain as
there are significant differences between various theoretical
models at the Neptune-mass regime. The equatorial declination of the
star makes it easily accessible to both Northern and Southern
ground-based facilities for follow-up observations.
\setcounter{footnote}{0}
\end{abstract}



\section{Introduction}
\label{sec:introduction}

Transiting exoplanets (TEPs) are tremendously useful objects for
studying the properties of planets outside of the solar system because
their photometric transits, combined with precise measurements of the
radial velocity variations of their host stars, enable determinations
of their masses and radii. Of the $\ga 90$ confirmed TEPs discovered
to date\footnote{e.g. http://exoplanet.eu}, all but five are Saturn or
Jupiter-size gas giant planets with masses above $0.1$\,\mjup. The
five TEPs below this limit, including the Super-Earths CoRoT-7b
\citep[$M = 0.015 \pm 0.003$\,\mjup, $R = 0.15 \pm
  0.008$\,\rjup;][]{leger:2009,queloz:2009}, and GJ~1214b \citep[$M =
  0.0206 \pm 0.0031$\,\mjup, $R = 0.239 \pm
  0.012$\,\rjup;][]{dc:2009}, and the Super-Neptunes GJ~436b \citep[$M
  = 0.078 \pm 0.006$\,\mjup, $R = 0.376 \pm
  0.022$\,\rjup;][]{butler:2004,gillon:2007,southworth:2009},
HAT-P-11b \citep[$M = 0.081 \pm 0.009$\,\mjup, $R = 0.422 \pm
  0.014$\,\rjup;][]{bakos:2010}, and Kepler-4b \citep[$M = 0.081 \pm
  0.014$\,\mjup, $R =
  0.515^{+0.2}_{-0.098}$\,\rjup;][]{borucki:2010,kipping:2010} are
likely composed primarily of elements heavier than hydrogren and
helium, and are therefore assumed to be qualitatively different from
the more massive gas giants. In addition to these five TEPs, the
candidate TEP Kepler-9d has an estimated radius of
$1.4$\,\rearth\ \citep{holman:2010}, and is most likely a low-mass
planet \citep{torres:2010}, but currently lacks a mass determination.

While the gas giant planets exhibit a wide range of radii at fixed
mass (and hence a great diversity in their physical structure at fixed
mass), the low mass TEPs, together with the six Solar System planets
smaller than Saturn, appear to follow a nearly monotonic relation
between mass and radius. The two Super-Neptunes with precise radius
measurements (GJ~436b and HAT-P-11b) have radii that are similar to
one another (to within 15\%) as well as to Uranus ($M =
0.0457$\,\mjup, $R = 0.358$\,\rjup\footnote{Solar system masses are
  taken from the IAU WG on NSFA report of current best estimates to
  the 2009 IAU General Assembly retrieved from
  http://maia.usno.navy.mil/NSFA/CBE.html; We adopt equatorial radii
  for the Solar System planets from \citet{seidelmann:2007}.}) and
Neptune ($M = 0.0540$\,\mjup, $R = 0.346$\,\rjup). While the mass and
radius of Kepler-4b given in the discovery paper
\citep[]{borucki:2010} are similar to those of GJ~436b and HAT-P-11b,
a reanalysis by \citet{kipping:2010} finds that the radius may be
$\sim 40$\% larger, though with a $20\%$ uncertainty, it may still be
similar to the other Super-Neptunes. The lack of significant scatter
in the radii among Uranus, Neptune, and the Super-Neptunes is perhaps
surprising given the vast range of radii permitted by theoretical
structure models for planets in this mass range. For example, the
theoretical models by \citet{fortney:2007} predict that a $1~{\rm
  Gyr}$ nonirradiated Neptune-mass planet may have a radius that
ranges from 0.14\,\rjup{} (pure iron composition) to 0.29\,\rjup{}
(pure water ice composition) or 0.86\,\rjup{} (pure gas
composition). These same models also predict that the radii of
gas-dominated Neptune-mass planets should be far more sensitive to
stellar irradiation than those of Jupiter-mass planets. For example, a
$1~{\rm Gyr}$ pure hydrogen-helium Neptune-mass planet at $0.045$\,AU
has a radius of $1.49$\,\rjup{} compared to $1.16$\,\rjup\ for a
similarly irradiated Jupiter-mass planet.

In this paper we present the discovery of \hatcurb{}, a TEP orbiting
the relatively bright star \hatcurCCgsc{} with a mass similar to that
of Neptune, but with a radius of $\hatcurPPrshort$\,\rjup\ or 65\%
larger than that of Neptune. This is the 26th TEP discovered by the
Hungarian-made Automated Telescope Network
\citep[HATNet;][]{bakos:2004} survey. In operation since 2003, HATNet
has now covered approximately 14\% of the sky, searching for TEPs
around bright stars ($8\lesssim r \lesssim 14.5$).  HATNet operates
six wide-field instruments: four at the Fred Lawrence Whipple
Observatory (FLWO) in Arizona, and two on the roof of the hangar
servicing the Smithsonian Astrophysical Observatory's Submillimeter
Array, in Hawaii.

The layout of the paper is as follows. In \refsecl{obs} we report the
detection of the photometric signal and the follow-up spectroscopic and
photometric observations of \hatcur{}.  In \refsecl{analysis} we
describe the analysis of the data, beginning with the determination of
the stellar parameters, continuing with a discussion of the methods
used to rule out nonplanetary, false positive scenarios which could
mimic the photometric and spectroscopic observations, and finishing
with a description of our global modeling of the photometry and radial
velocities.  Our findings are discussed in \refsecl{discussion}.

\section{Observations}
\label{sec:obs}

\subsection{Photometric detection}
\label{sec:detection}

The transits of \hatcurb{} were detected with the HAT-5 and HAT-6
telescopes in Arizona, and with the HAT-8 and HAT-9 telescopes in
Hawaii.  Two regions around \hatcurCCgsc{}, corresponding to fields
internally labeled as \hatcurfieldA\ and \hatcurfieldB, were both
observed on a nightly basis between 2009 Jan and 2009 Aug, whenever
weather conditions permitted.  For field \hatcurfieldA\ we gathered
11,500 exposures of 5 minutes at a 5.5 minute cadence.  Approximately
1500 of these exposures were rejected by our photometric pipeline
because they yielded poor photometry for a significant number of
stars. Each image contained approximately 17,000 stars down to Sloan
$r \sim 14.5$. For the brightest stars in the field, we achieved a
per-image photometric precision of 4\,mmag. For field
\hatcurfieldB\ we gathered 5200 exposures with the same exposure time
and cadence; we rejected aproximately 700 exposures. Each image
contained approximately 19,000 stars down to Sloan $r \sim 14.5$. We
achieved a similar photometric precision for the brightest stars in
this field.

\begin{figure}[!ht]
\plotone{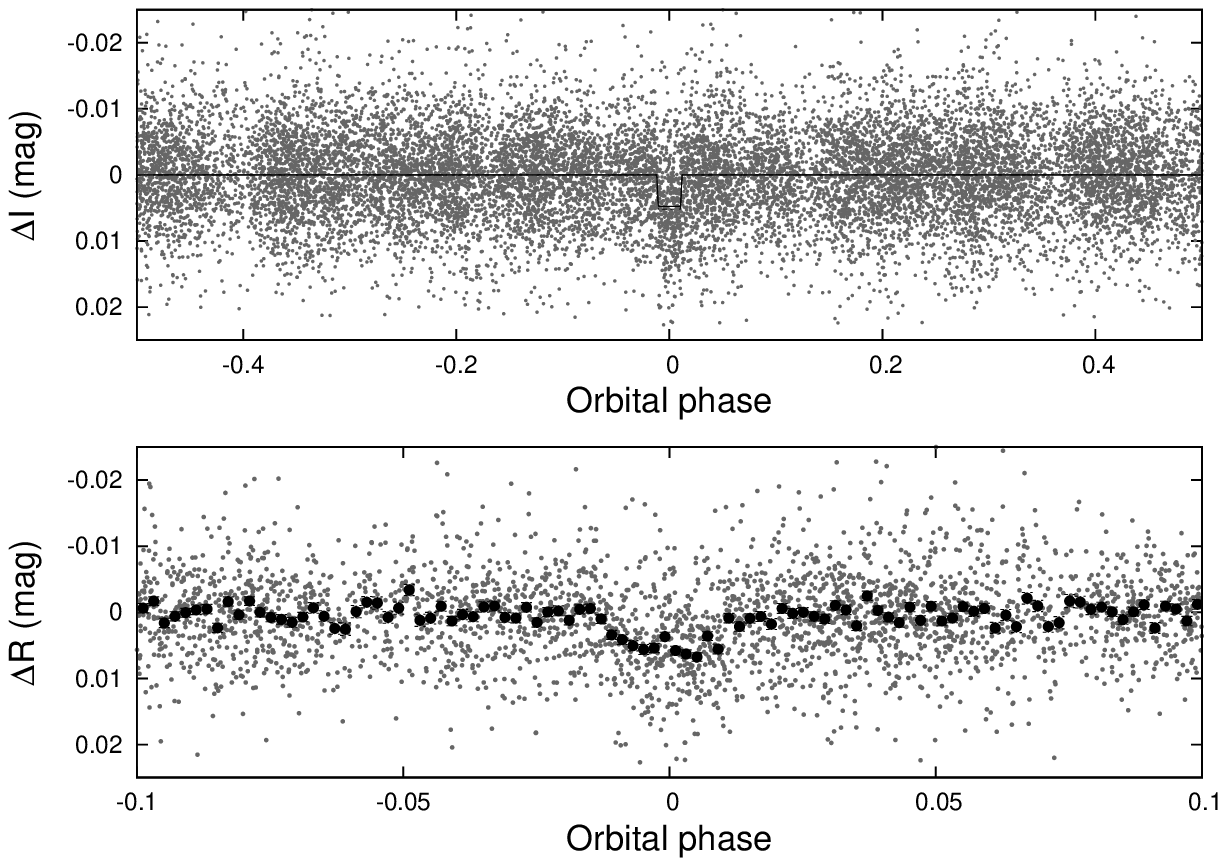}
\caption{
	Unbinned \lc{} of \hatcur{} including all 14,500 instrumental
        Sloan \band{r} 5.5 minute cadence measurements obtained with
        the HAT-5, HAT-6, HAT-8 and HAT-9 telescopes of HATNet (see
        the text for details), and folded with the period $P =
        \hatcurLCPprec$\,days resulting from the global fit described
        in \refsecl{analysis}).  The solid line shows the ``P1P3''
        transit model fit to the light curve (\refsecl{globmod}). The
        lower panel shows a zoomed-in view of the transit; the dark
        filled points show the light curve binned in phase using a
        bin-size of 0.002.
\label{fig:hatnet}}
\end{figure}

The calibration of the HATNet frames was carried out using standard
photometric procedures.  The calibrated images were then subjected to
star detection and astrometry, as described in \cite{pal:2006}.
Aperture photometry was performed on each image at the stellar
centroids derived from the Two Micron All Sky Survey
\citep[2MASS;][]{skrutskie:2006} catalog and the individual
astrometric solutions.  The resulting \lcs\ were decorrelated (cleaned
of trends) using the External Parameter Decorrelation \citep[EPD;
  see][]{bakos:2010} technique in ``constant'' mode and the Trend
Filtering Algorithm \citep[TFA; see][]{kovacs:2005}.  The \lcs{} were
searched for periodic box-shaped signals using the Box Least-Squares
\citep[BLS; see][]{kovacs:2002} method.  We detected a significant
signal in the \lc{} of \hatcurCCgsc{} (also known as
\hatcurCCtwomass{}; $\alpha = \hatcurCCra$, $\delta = \hatcurCCdec$;
J2000; V=\hatcurCCtassmv\ \citealp{droege:2006}), with an apparent
depth of $\sim\hatcurLCdip$\,mmag ($\sim 5.5$\,mmag when using TFA in
signal-reconstruction mode), and a period of $P=\hatcurLCPshort$\,days
(see \reffigl{hatnet}).  The drop in brightness had a
first-to-last-contact duration, relative to the total period, of $q =
\hatcurLCq$, corresponding to a total duration of $Pq =
\hatcurLCdurhr$~hr.

We removed the transits from the combined
\hatcurfieldA/\hatcurfieldB\ light curve and searched for additional
transiting objects using the BLS method; no significant signals were
found in the data. We also searched the transit-cleaned light curve
for periodic variations (e.g.~due to stellar rotation) using the
Discrete Fourier Transform \citep[e.g.][]{kurtz:1985}, and found no
coherent variation with an amplitude greater than $0.4$\,mmag.

\subsection{Reconnaissance Spectroscopy}
\label{sec:recspec}

As is routine in the HATNet project, all candidates are subjected to
careful scrutiny before investing valuable time on large telescopes. 
This includes spectroscopic observations at relatively modest
facilities to establish whether the transit-like feature in the light
curve of a candidate might be due to astrophysical phenomena other than
a planet transiting a star.  Many of these false positives are
associated with large radial-velocity variations in the star (tens of
\kms) that are easily recognized.

To carry out this reconnaissance spectroscopy, we made use of the
Tillinghast Reflector Echelle Spectrograph
\citep[TRES;][]{furesz:2008} on the 1.5\,m Tillinghast Reflector at
FLWO. This instrument provides high-resolution spectra which, with
even modest signal-to-noise (S/N) ratios, are suitable for deriving
RVs with moderate precision ($\la 0.3$\,\kms) for slowly rotating
stars. We also use these spectra to estimate effective temperatures,
surface gravities, and projected rotational velocities of the target
star. Using the medium fiber on TRES, we obtained two spectra of
\hatcur\ on the nights of 2009 Dec 26 and 2009 Dec 27. The spectra
have a resolution of $\lambda/\Delta\lambda \approx 44,\!000$ and a
wavelength coverage of $\sim$\,3900-8900\,\AA. The spectra were
extracted and analyzed according to the procedure outlined by
\cite{buchhave:2010} and \cite{quinn:2010}. The individual velocity
measurements of $14.62$\,\kms\ and $14.81$\,\kms\ were consistent with
no detectable RV variation within the measurement precision.  Both
spectra were single-lined, i.e., there is no evidence for additional
stars in the system.  The atmospheric parameters we infer from these
observations are the following: effective temperature $\teffstar =
\hatcurTRESteff\,K$, surface gravity $\loggstar = \hatcurTRESlogg$
(log cgs), and projected rotational velocity $\vsini =
\hatcurTRESvsini\,\kms$.  The effective temperature corresponds to an
early K dwarf.  The mean heliocentric RV of \hatcur\ is $\gamma_{\rm
  RV} = \hatcurTRESgamma$\,\kms.


\subsection{High resolution, high S/N spectroscopy}
\label{sec:hispec}

Given the significant transit detection by HATNet, and the encouraging
TRES results that rule out obvious false positives, we proceeded with
the follow-up of this candidate by obtaining high-resolution, high-S/N
spectra to characterize the RV variations, and to refine the
determination of the stellar parameters.  For this we used the HIRES
instrument \citep{vogt:1994} on the Keck~I telescope located on Mauna
Kea, Hawaii, between 2009 Dec and 2010 June.  The width of the
spectrometer slit was $0\farcs86$, resulting in a resolving power of
$\lambda/\Delta\lambda \approx 55,\!000$, with a wavelength coverage
of $\sim$3800--8000\,\AA\@.

\begin{figure} [ht]
\plotone{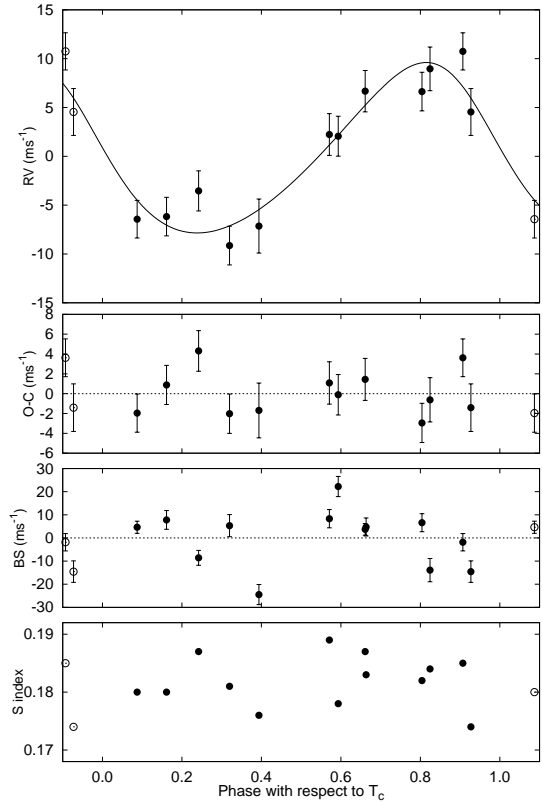}
\caption{
	{\em Top panel:} Keck/HIRES RV measurements for \hbox{\hatcur{}}
    shown as a function of orbital phase, along with our best-fit model
    (see \reftabl{planetparam}).  Zero phase corresponds to the time of
    mid-transit.  The center-of-mass velocity has been subtracted.
	{\em Second panel:} Velocity $O\!-\!C$ residuals from the best fit. 
    The error bars include a component from astrophysical/instrumental jitter
    ($\hatcurRVjitter$\,\ms) added in quadrature to the formal errors
    (see \refsecl{globmod}).
	{\em Third panel:} Bisector spans (BS), with the mean value
    subtracted.  The measurement from the template spectrum is included
    (see \refsecl{blend}).
	{\em Bottom panel:} Chromospheric activity index $S_{\rm HK}$
    measured from the Keck spectra.
	Note the different vertical scales of the panels. Observations
    shown twice are represented with open symbols.
\label{fig:rvbis}}
\end{figure}

We obtained 12 exposures through an iodine gas absorption cell, which
was used to superimpose a dense forest of $\mathrm{I}_2$ lines on the
stellar spectrum and establish an accurate wavelength fiducial
\citep[see][]{marcy:1992}.  An additional exposure was taken
without the iodine cell, for use as a template in the reductions. 
Relative RVs in the solar system barycentric frame were derived as
described by \cite{butler:1996}, incorporating full modeling of the
spatial and temporal variations of the instrumental profile.  The RV
measurements and their uncertainties are listed in \reftabl{rvs}.  The
period-folded data, along with our best fit described below in
\refsecl{analysis}, are displayed in \reffigl{rvbis}.

\begin{figure} [ht]
\plotone{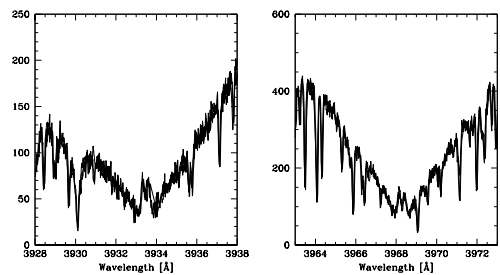}
\caption{
  Calcium K (left) and H (right) line profile in selected Keck/HIRES
  observations of \hatcur. Both panels show three spectra overlaid;
  data taken at high, median, and low activity, as characterized by
  the $S_{\rm HK}$ index. Low-level emission is clearly detected in
  the cores of both lines. The lack of apparent variation (the three
  plotted spectra are indistinguishable) indicates the chromospheric
  stability of this star over the course of our observations. The
  spectra are matched to a common flux/wavelength scale using points
  outside the H and K line cores. The vertical axes give the counts in
  units of CCD $e^{-}$ per wavelength bin for the reference spectrum.
\label{fig:CaHK}}
\end{figure}

In the same figure we show also the $S_{\rm HK}$ index, which is a
measure of the chromospheric activity of the star derived from the
flux in the cores of the \ion{Ca}{2} H and K lines (\reffigl{CaHK}
shows representative Keck spectra including the H and K lines for
\hatcur).  This index was computed and calibrated to the scale of
\citet{vaughan:1978} following the procedure described by
\citet{isaacson:2010}. We find a median value of $S_{\rm HK} =
\hatcurRHKsmed$ with a standard deviation of \hatcurRHKsrms. Assuming
$B-V = \hatcurRHKbv$ based on the effective temperature measured in
\refsecl{stelparam}, this corresponds to $\log R^{\prime}_{\rm HK} =
\hatcurRHKrhkmed$ \citep{noyes:1984}. From \citet{isaacson:2010} the
lower tenth percentile $S_{\rm HK}$ value among California Planet
Search (CPS) targets with $B-V = \hatcurRHKbv$ is
$\hatcurRHKsbaselevel$. The measured $S_{\rm HK}$ value is only
slightly higher than this, implying that \hatcur\ is a
chromospherically quiet star. We do not detect any significant
variation of the $S_{\rm HK}$ index correlated with orbital phase;
such a correlation might have indicated that the RV variations could
be due to stellar activity, casting doubt on the planetary nature of
the candidate.

\ifthenelse{\boolean{emulateapj}}{
    \begin{deluxetable*}{lrrrrrr}
}{
    \begin{deluxetable}{lrrrrrr}
}
\tablewidth{0pc}
\tablecaption{
	Relative radial velocities, bisector spans, and activity index
	measurements of \hatcur{}. 
	\label{tab:rvs}
}
\tablehead{
	\colhead{BJD} & 
	\colhead{RV\tablenotemark{a}} & 
	\colhead{\ensuremath{\sigma_{\rm RV}}\tablenotemark{b}} & 
	\colhead{BS} & 
	\colhead{\ensuremath{\sigma_{\rm BS}}} & 
	\colhead{\ensuremath{S_{\rm HK}}\tablenotemark{c}} &
        \colhead{Phase} \\
	\colhead{\hbox{(2,454,000$+$)}} & 
	\colhead{(\ms)} & 
	\colhead{(\ms)} &
	\colhead{(\ms)} &
    \colhead{(\ms)} &
	\colhead{} &
        \colhead{}
}
\startdata
$ 1193.11925 $ & $     6.67 $ & $     1.87 $ & $     3.71 $ & $     2.51 $ & $    0.1870 $ &$   0.661 $\\
$ 1193.12855 $ & \nodata      & \nodata      & $     4.74 $ & $     3.90 $ & $    0.1830 $ &$   0.663 $\\
$ 1194.16006 $ & $    10.75 $ & $     1.62 $ & $    -1.85 $ & $     3.76 $ & $    0.1850 $ &$   0.907 $\\
$ 1252.02017 $ & $     2.23 $ & $     1.89 $ & $     8.33 $ & $     3.95 $ & $    0.1890 $ &$   0.571 $\\
$ 1285.14629 $ & $    -7.14 $ & $     2.58 $ & $   -24.47 $ & $     4.35 $ & $    0.1760 $ &$   0.394 $\\
$ 1320.84584 $ & $     8.96 $ & $     2.00 $ & $   -13.90 $ & $     5.02 $ & $    0.1840 $ &$   0.824 $\\
$ 1343.78649 $ & $    -3.54 $ & $     1.79 $ & $    -8.59 $ & $     3.23 $ & $    0.1870 $ &$   0.242 $\\
$ 1350.92272 $ & $     4.54 $ & $     2.18 $ & $   -14.55 $ & $     4.65 $ & $    0.1740 $ &$   0.927 $\\
$ 1351.91410 $ & $    -6.17 $ & $     1.70 $ & $     7.80 $ & $     4.04 $ & $    0.1800 $ &$   0.161 $\\
$ 1372.77275 $ & $    -6.44 $ & $     1.65 $ & $     4.63 $ & $     2.63 $ & $    0.1800 $ &$   0.087 $\\
$ 1373.75770 $ & $    -9.14 $ & $     1.71 $ & $     5.32 $ & $     4.79 $ & $    0.1810 $ &$   0.320 $\\
$ 1374.91646 $ & $     2.06 $ & $     1.78 $ & $    22.23 $ & $     4.36 $ & $    0.1780 $ &$   0.593 $\\
$ 1375.80851 $ & $     6.63 $ & $     1.71 $ & $     6.59 $ & $     3.90 $ & $    0.1820 $ &$   0.804 $\\
[-1.5ex]
\enddata
\tablenotetext{a}{
	The zero-point of these velocities is arbitrary. An overall offset
    $\gamma_{\rm rel}$ fitted to these velocities in \refsecl{globmod}
    has {\em not} been subtracted.
}
\tablenotetext{b}{
	Internal errors excluding the component of astrophysical/instrumental jitter
    considered in \refsecl{globmod}.
}
\tablenotetext{c}{
	$S_{\rm HK}$ chromospheric activity index, calibrated to the
	scale of \citet{vaughan:1978}.
}
\ifthenelse{\boolean{rvtablelong}}{
	\tablecomments{
             For the iodine-free template exposures there is no RV
             measurement, but the BS and $S_{\rm HK}$ index can still
             be determined.
	}
}{
    \tablecomments{
		For the iodine-free template exposures there is no RV
		measurement, but the BS and S index can still be determined.
		This table is presented in its entirety in the
		electronic edition of the Astrophysical Journal.  A portion is
		shown here for guidance regarding its form and content.
	}
} 
\ifthenelse{\boolean{emulateapj}}{
    \end{deluxetable*}
}{
    \end{deluxetable}
}

\subsection{Photometric follow-up observations}
\label{sec:phot}

\begin{figure}[!ht]
\plotone{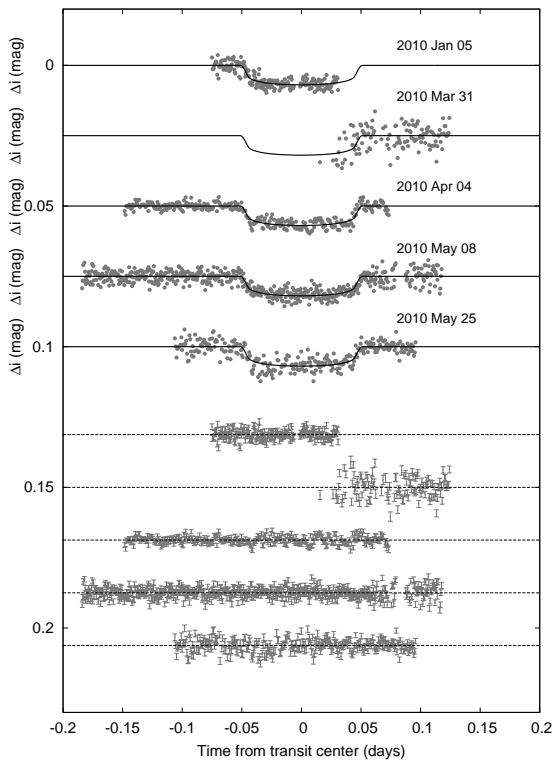}
\caption{
  Unbinned instrumental Sloan \band{i} light curves, acquired with
  KeplerCam at the \flwof{} telescope.  The light curves have been
  EPD- and TFA-processed, as described in \refsec{globmod}.
    The dates of the events are indicated.  Curves after the first are
    displaced vertically for clarity.  Our best fit from the global
    modeling described in \refsecl{globmod} is shown by the solid
    lines.  Residuals from the fits are displayed at the bottom, in the
    same order as the top curves.  The error bars represent the photon
    and background shot noise, plus the readout noise.
\label{fig:lc}}
\end{figure}

\begin{figure}[!ht]
\plotone{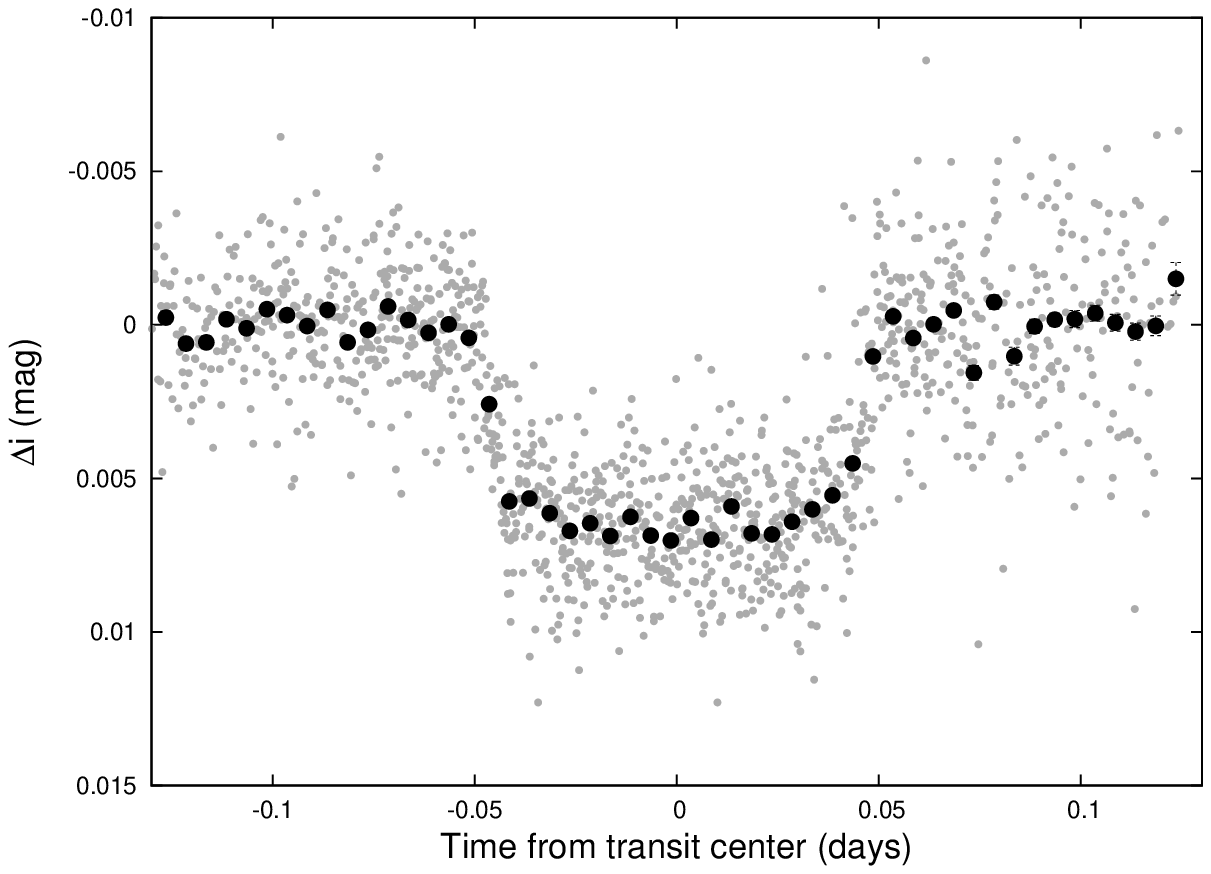}
\caption{
  The combined Sloan \band{i} KeplerCam \lc\ of \hatcur\ folded with
  the period $P = \hatcurLCPprec$\,days resulting from the global fit
  described in \refsecl{analysis}. The dark filled circles show the
  \lc\ binned in folded time with a bin-size of 0.005\,days. The
  median uncertainty on the binned points is $0.19$\,mmag.
\label{fig:phfu}}
\end{figure}

In order to permit a more accurate modeling of the light curve, we
conducted additional photometric observations with the KeplerCam CCD
camera on the \flwof{} telescope.  We observed five transit events of
\hatcur{} on the nights of 2010 Jan 05, Mar 31, Apr 04, May 08 and May 25 (\reffigl{lc}).  
These observations are summarized in \reftabl{phfusummary}.

\ifthenelse{\boolean{emulateapj}}{
    \begin{deluxetable*}{llrrr}
}{
    \begin{deluxetable}{llrrr}
}
\tablewidth{0pc}
\tabletypesize{\scriptsize}
\tablecaption{
    Summary of photometric follow-up observations
    \label{tab:phfusummary}
}
\tablehead{
    \colhead{Facility}  &
    \colhead{Date} &
    \colhead{Number of Images} &
    \colhead{Median Cadence (s)} &
    \colhead{Filter}
}
\startdata
KeplerCam/\flwof{} & 2010 Jan 05 & 191 & 40 & Sloan \band{i} \\
KeplerCam/\flwof{} & 2010 Mar 31 & 161 & 59 & Sloan \band{i} \\
KeplerCam/\flwof{} & 2010 Apr 04 & 291 & 64 & Sloan \band{i} \\
KeplerCam/\flwof{} & 2010 May 08 & 596 & 44 & Sloan \band{i} \\
KeplerCam/\flwof{} & 2010 May 25 & 298 & 59 & Sloan \band{i} \\
[-1.5ex]
\enddata
\ifthenelse{\boolean{emulateapj}}{
    \end{deluxetable*}
}{
    \end{deluxetable}
}

The reduction of these images, including basic calibration,
astrometry, and aperture photometry, was performed as described by
\citet{bakos:2010}.  We performed EPD and TFA to remove trends
simultaneously with the light curve modeling (for more details, see
\refsecl{analysis}, and \citealt{bakos:2010}).  The final time series
are shown in the top portion of \reffigl{lc}, along with our best-fit
transit \lc{} model described below; the individual measurements are
reported in \reftabl{phfu}. The combined phase-folded follow-up light
curve is displayed in \reffigl{phfu}.


\begin{deluxetable}{lrrrr}
\tablewidth{0pc}
\tablecaption{High-precision differential photometry of 
	\hatcur\label{tab:phfu}}
\tablehead{
	\colhead{BJD} & 
	\colhead{Mag\tablenotemark{a}} & 
	\colhead{\ensuremath{\sigma_{\rm Mag}}} &
	\colhead{Mag(orig)\tablenotemark{b}} & 
	\colhead{Filter} \\
	\colhead{\hbox{~~~~(2,400,000$+$)~~~~}} & 
	\colhead{} & 
	\colhead{} &
	\colhead{} & 
	\colhead{}
}
\startdata
$ 55202.94802 $ & $  -0.00279 $ & $   0.00105 $ & $  10.17790 $ & $ i$\\
$ 55202.94847 $ & $   0.00137 $ & $   0.00106 $ & $  10.18150 $ & $ i$\\
$ 55202.94909 $ & $   0.00132 $ & $   0.00106 $ & $  10.18130 $ & $ i$\\
$ 55202.94954 $ & $   0.00079 $ & $   0.00106 $ & $  10.18110 $ & $ i$\\
$ 55202.95018 $ & $  -0.00047 $ & $   0.00105 $ & $  10.17920 $ & $ i$\\
$ 55202.95063 $ & $  -0.00029 $ & $   0.00106 $ & $  10.17980 $ & $ i$\\
$ 55202.95126 $ & $  -0.00131 $ & $   0.00106 $ & $  10.17960 $ & $ i$\\
$ 55202.95171 $ & $  -0.00097 $ & $   0.00105 $ & $  10.17950 $ & $ i$\\
$ 55202.95237 $ & $   0.00334 $ & $   0.00107 $ & $  10.18530 $ & $ i$\\
$ 55202.95282 $ & $  -0.00366 $ & $   0.00106 $ & $  10.17800 $ & $ i$\\
[-1.5ex]
\enddata
\tablenotetext{a}{
	The out-of-transit level has been subtracted. These magnitudes have
	been subjected to the EPD and TFA procedures, carried out
	simultaneously with the transit fit.
}
\tablenotetext{b}{
	Raw magnitude values without application of the EPD and TFA
	procedures.
}
\tablecomments{
    This table is available in a machine-readable form in the online
    journal.  A portion is shown here for guidance regarding its form
    and content.
}
\end{deluxetable}

\section{Analysis}
\label{sec:analysis}

\subsection{Properties of the parent star}
\label{sec:stelparam}

Fundamental parameters of the host star \hatcur{} such as the mass
(\mstar) and radius (\rstar), which are needed to infer the planetary
properties, depend strongly on other stellar quantities that can be
derived spectroscopically.  For this we have relied on our template
spectrum obtained with the Keck/HIRES instrument, and the analysis
package known as Spectroscopy Made Easy \citep[SME;][]{valenti:1996},
along with the atomic line database of \cite{valenti:2005}.  SME
yielded the following values and uncertainties:
effective temperature $\teffstar=\hatcurSMEiteff$\,K, 
stellar surface gravity $\loggstar=\hatcurSMEilogg$\,(cgs),
metallicity $\feh=\hatcurSMEizfeh$\,dex, and 
projected rotational velocity $\vsini=\hatcurSMEivsin\,\kms$, 
in which the uncertainties for $\teffstar$ and $\feh$ have been
increased by a factor of two over their formal values to include our
estimates of the systematic uncertainties.

In principle the effective temperature and metallicity, along with the
surface gravity taken as a luminosity indicator, could be used as
constraints to infer the stellar mass and radius by comparison with
stellar evolution models.  However, the effect of \loggstar\ on the
spectral line shapes is quite subtle, and as a result it is typically
difficult to determine accurately, so that it is a rather poor
luminosity indicator in practice.  For planetary transits a stronger
constraint is often provided by \rhostar\ the mean stellar density,
which is closely related to the \arstar\ normalized semimajor axis.
The quantity \rhostar\ can be derived directly from the combination of
the transit \lcs\ \citep{seager:2003,sozzetti:2007} and the RV data
(required for eccentric cases, see \refsecl{globmod}).  This, in turn,
allows us to improve on the determination of the spectroscopic
parameters by supplying an indirect constraint on the weakly
determined spectroscopic value of \loggstar\ that removes
degeneracies.  We take this approach here, as described below.  The
validity of our assumption, namely that the adequate physical model
describing our data is a planetary transit (as opposed to a blend), is
shown later in \refsecl{blend}.

Our values of \teffstar, \loggstar, and \feh\ were used to determine
auxiliary quantities needed in the global modeling of the follow-up
photometry and radial velocities (specifically, the limb-darkening
coefficients).  This modeling, the details of which are described in
\refsecl{globmod}, uses a Monte Carlo approach to deliver the
numerical probability distribution of \rhostar\ and other fitted
variables.  For further details we refer the reader to
\cite{pal:2009b}.  When combining \rhostar\ (a luminosity proxy) with
assumed Gaussian distributions for \teffstar\ and \feh\ based on the
SME determinations, a comparison with stellar evolution models allows
the probability distributions of other stellar properties to be
inferred, including \loggstar.  Here we use the stellar evolution
calculations from the \hatcurisofull\ series by \cite{\hatcurisocite}.

\begin{figure}[!ht]
\plotone{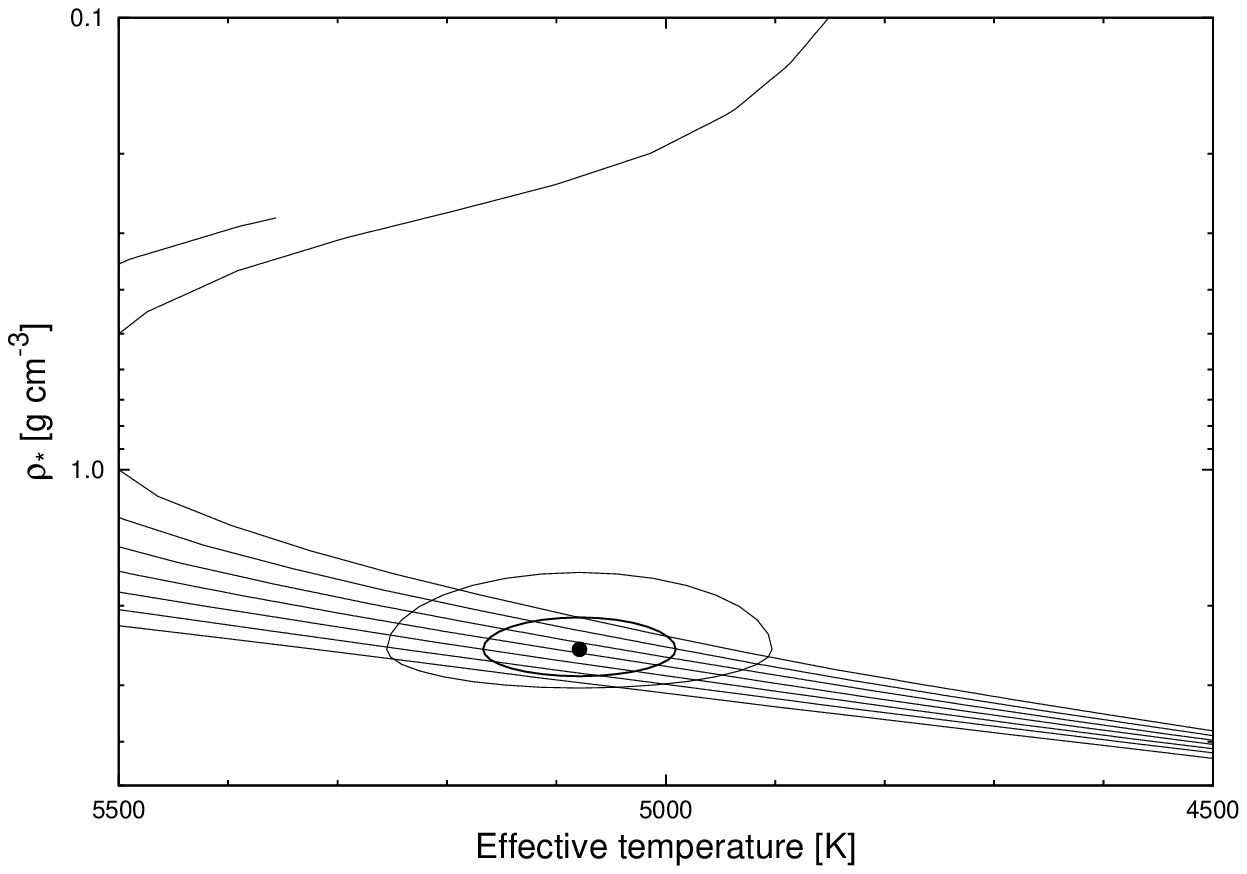}
\plotone{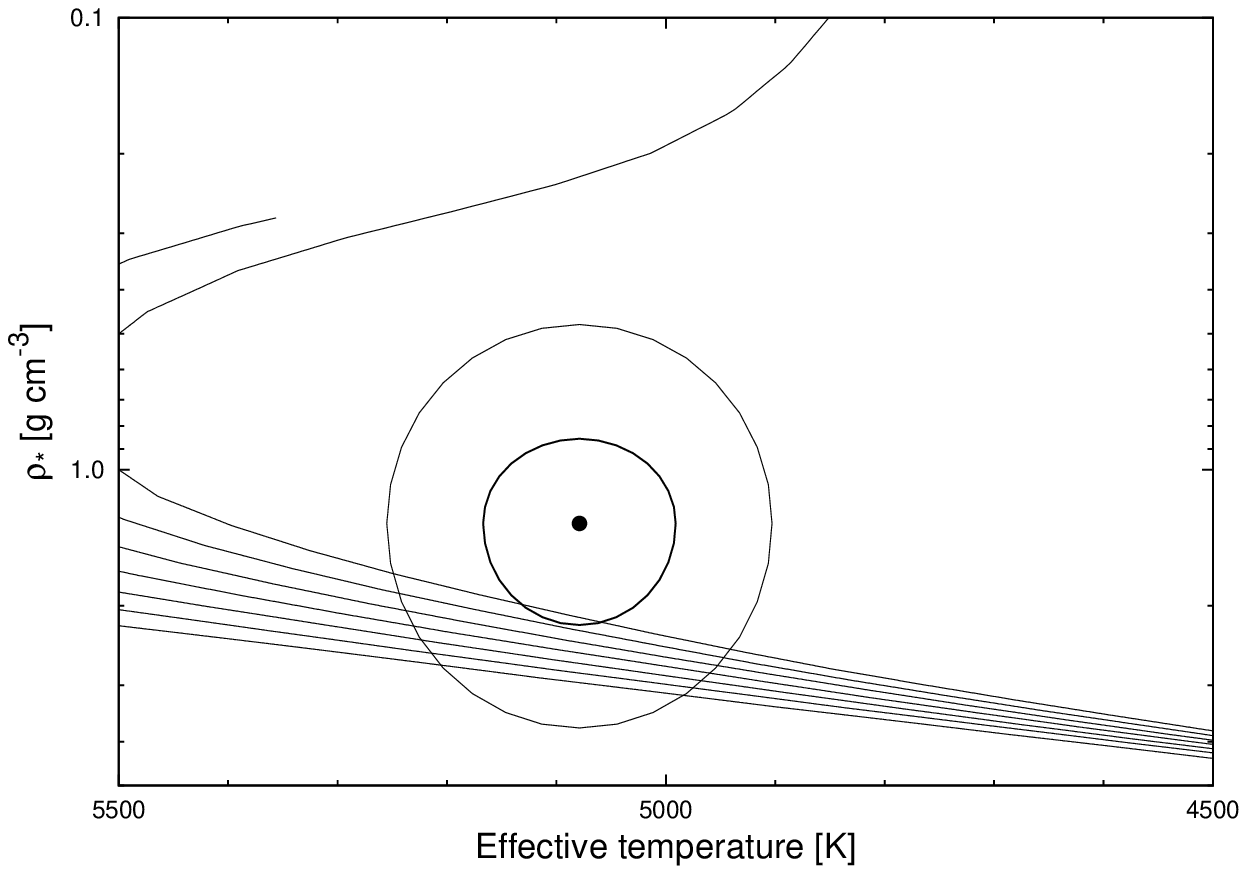}
\plotone{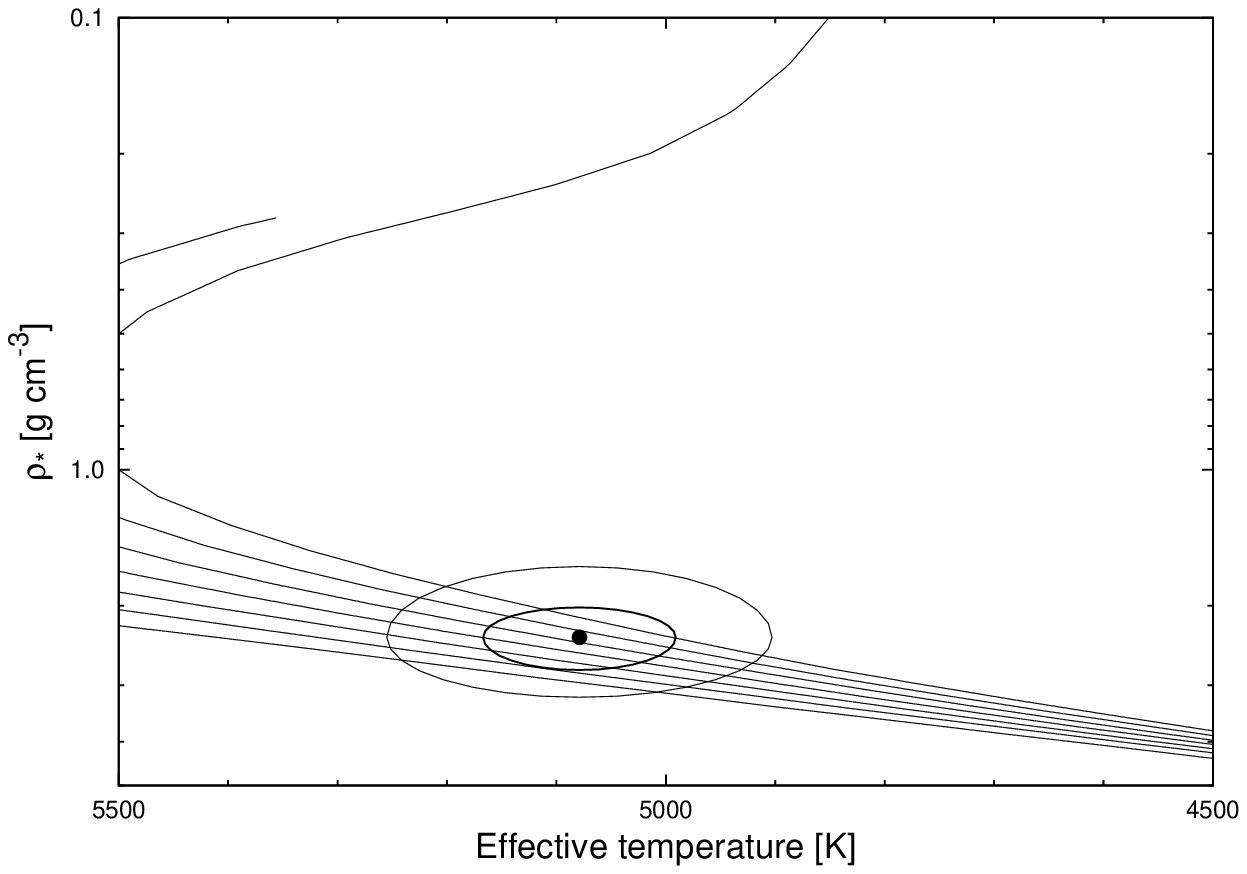}
\caption{
  Model isochrones from \cite{\hatcurisocite} for the measured
  metallicity of \hatcur, \feh = \hatcurSMEiizfehshort, and ages in
  2\,Gyr increments between 1 and 13\,Gyr (left to right).  Note that
  a logarithmic scale is used for the vertical axis. The measured
  values of $\teffstar$ and $\rhostar$ are shown together with their
  approximate $1\sigma$ and $2\sigma$ confidence boundaries for models
  where the eccentricity is fixed to 0 (top), the eccentricity is
  allowed to vary (bottom), and the eccentricity is allowed to vary,
  but only models which match to an isochrone with an age between
  0.1\,Gyr and 13.8\,Gyr are accepted (bottom). We adopt the
  parameters shown in the bottom panel.
\label{fig:iso}}
\end{figure}

For the case of \hatcurb, the eccentricity is poorly constrained
by the RV data due to the low semiamplitude of the signal. This in
turn leads to a significant uncertainty on \rhostar. However, not all
combinations of \feh, \teffstar\ and \rhostar\ are realized by
physical stellar models. In particular, if we conservatively adopt a
maximum stellar age of $13.8$\,Gyr, corresponding approximately to the
age of the universe \citep[][find $13.75 \pm
  0.11$\,Gyr]{komatsu:2010}, and a minimum age of $100$\,Myr,
corresponding roughly to the zero-age main-sequence\footnote{The lack
  of evidence for stellar activity indicates that \hatcur\ is unlikely
  to be a pre-main sequence star.}, stars with $\teffstar =
\hatcurSMEiteffshort$ and $\feh = \hatcurSMEizfehshort$ are not found
to have densities in the range $0.24\,\gcmc < \rhostar < 2.06\,\gcmc$
or surface gravities in the range $3.915 < \loggstar < 4.514$ (here
$0.24\,\gcmc$ corresponds to an evolved star with $M = 0.94$\,\msun,
while $2.06\,\gcmc$ corresponds to a main sequence star with $M =
0.79$\,\msun). \reffigl{iso} shows the inferred location of the star
in a diagram of \rhostar\ versus \teffstar, analoguous to the
classical H-R diagram, for three cases: fixing the eccentricity of the
orbit to zero, allowing the eccentricity to vary, and allowing the
eccentricity to vary, but only accepting parameter combinations which
match to a position in the YY isochrones with $0.1\,{\rm Gyr} < {\rm
  age} < 13.8\,{\rm Gyr}$. The stellar properties and their
approximate 1$\sigma$ and 2$\sigma$ confidence boundaries are
displayed against the backdrop of \cite{\hatcurisocite} isochrones for
the measured metallicity of \feh\ = \hatcurSMEizfehshort, and a range
of ages. For the zero eccentricity case the comparison against the
model isochrones was carried out for each of 30,000 Monte Carlo trial
sets (see \refsecl{globmod}). We find good overlap between the trials
and the model isochrones--in 71\% of the trials, the \feh,
\teffstar\ and \rhostar\ parameter combination matched to a physical
location in the \hbox{H-R} diagram that has an age that is within the
aforementioned range. However, when the eccentricity is allowed to
vary, the model for the light curves and RV data tends toward low
values of $\rhostar$ which may only be fit by pre-main sequence
stellar models, or stellar models older than the age of the
universe. In this case only 15\% out of 100,000 Monte Carlo trial sets
match to physical locations in the \hbox{H-R} diagram with ages within
the allowed range. By requiring the star to have an age between
0.1\,Gyr and 13.8\,Gyr, we effectively impose a tighter constraint on
the orbital eccentricity than is possible from the RV data alone (we
find an eccentricity of $e = \hatcurRVeccen$, as compared with $e =
0.24 \pm 0.12$ when not requiring a match to the stellar models; see
also \refsecl{globmod}).

Adopting the parameters which result from allowing the eccentricity to
vary while requiring the star to have an age between 0.1\,Gyr and
13.8\,Gyr yields a stellar surface gravity of $\loggstar =
\hatcurISOlogg$, which is very close to the value from our SME
analysis. The values for the atmospheric parameters of the star are
collected in \reftabl{stellar}, together with the adopted values for
the macroturbulent and microturbulent velocities.


The stellar evolution modeling provides color indices that may be
compared against the measured values as a sanity check. The best
available measurements are the near-infrared magnitudes from the 2MASS
Catalogue \citep{skrutskie:2006},
$J_{\rm 2MASS}=\hatcurCCtwomassJmag$, 
$H_{\rm 2MASS}=\hatcurCCtwomassHmag$ and 
$K_{\rm 2MASS}=\hatcurCCtwomassKmag$;
which we have converted to the photometric system of the models (ESO)
using the transformations by \citet{carpenter:2001}.  The resulting
measured color index is $J-K = \hatcurCCesoJKmag$.  This is within
1$\sigma$ of the predicted value from the isochrones of $J-K =
\hatcurISOJK$.  The distance to the object may be computed from the
absolute $K$ magnitude from the models ($M_{\rm K}=\hatcurISOMK$) and
the 2MASS $K_s$ magnitude, which has the advantage of being less
affected by extinction than optical magnitudes.  The result is
$\hatcurXdist$\,pc, where the uncertainty excludes possible
systematics in the model isochrones that are difficult to quantify.

An additional check on our stellar model can be performed by comparing
the isochrone-based age estimate to activity-based age estimates.
Using the activity-rotation and activity-age relations from
\citet[][equations 5 and 3]{mamajek:2008} we convert the value of
$\log R^{\prime}_{\rm HK}$ determined in \refsecl{hispec} into a
Rossby number ($R_{O} = P_{\rm rot}/\tau_c$, where $P_{\rm rot}$ is
the rotation period and $\tau_c$ is the convective turnover
time-scale) and an age. We find $R_{O} = \hatcurRHKro$, and
$\log(\tau) = \hatcurRHKlogage$ or $\tau = \hatcurRHKage$\,Gyr, where
we adopt the estimated uncertainties on $R_{O}$ and $\log(\tau)$ from
\citet{mamajek:2008}. The $R_{O}$ value may be converted to a rotation
period using the relation for $\tau_c$ given by \citet{noyes:1984}. We
find $P_{\rm rot} = \hatcurRHKprot$\,d. The rotation period and color
may also be used to obtain a separate age estimate from the
gyrochronology relation given by \citet[][equation
  12]{mamajek:2008}. This gives $\tau = \hatcurRHKgyroage$\,Gyr, where
the scatter on this relation is less than the scatter on the
age-activity relation because it includes a correction for stellar
color. The age inferred from $\log R^{\prime}_{\rm HK}$ is consistent
with the isochrone-based age of $\hatcurISOage$\,Gyr. The equatorial
rotation period inferred from the spectroscopic determination of $v
\sin i$ assuming $\sin i = 1$ is $P_{\rm rot,v\sin i} =
\hatcurRHKprotvsini$\,d, which is shorter than the expected value
based on $\log R^{\prime}_{\rm HK}$, though the upper limit is poorly
constrained.

As discussed below in \refsecl{globmod} we measure a RV jitter of
$\hatcurRVjitter$\,\ms\ for \hatcur. From \citet{isaacson:2010} the
lower tenth percentile jitter among HIRES/Keck observations for CPS
stars with $0.7 < B-V < 1.0$ and $S_{\rm HK} = \hatcurRHKsmed$ is
$2.17$\,\ms, implying that \hatcur\ has an exceptionally low jitter
value--only a handful of stars in this color range have measured
jitter values less than that of \hatcur. We note that the jitter may
be higher ($2.4$\,\ms) if the orbit is circular, though this value is
still quite low.

\begin{deluxetable}{lrl}
\tablewidth{0pc}
\tabletypesize{\scriptsize}
\tablecaption{
	Stellar parameters for \hatcur{}
	\label{tab:stellar}
}
\tablehead{
	\colhead{~~~~~~~~Parameter~~~~~~~~}	&
	\colhead{Value} &
	\colhead{Source}
}
\startdata
\noalign{\vskip -3pt}
\sidehead{Spectroscopic properties}
~~~~$\teffstar$ (K)\dotfill         &  \hatcurSMEteff       & SME\tablenotemark{a}\\
~~~~$\feh$\dotfill                  &  \hatcurSMEzfeh       & SME                 \\
~~~~$\vsini$ (\kms)\dotfill         &  \hatcurSMEvsin       & SME                 \\
~~~~$\vmac$ (\kms)\dotfill          &  \hatcurSMEvmac       & SME                 \\
~~~~$\vmic$ (\kms)\dotfill          &  \hatcurSMEvmic       & SME                 \\
~~~~$\gamma_{\rm RV}$ (\kms)\dotfill&  \hatcurTRESgamma      & TRES                  \\
\sidehead{Photometric properties}
~~~~$V$ (mag)\dotfill               &  \hatcurCCtassmv      & TASS                \\
~~~~$V\!-\!I_C$ (mag)\dotfill       &  \hatcurCCtassvi      & TASS                \\
~~~~$J$ (mag)\dotfill               &  \hatcurCCtwomassJmag & 2MASS           \\
~~~~$H$ (mag)\dotfill               &  \hatcurCCtwomassHmag & 2MASS           \\
~~~~$K_s$ (mag)\dotfill             &  \hatcurCCtwomassKmag & 2MASS           \\
\sidehead{Derived properties}
~~~~$\mstar$ ($\msun$)\dotfill      &  \hatcurISOmlong      & \hatcurisoshort+\hatcurlumind+SME \tablenotemark{b}\\
~~~~$\rstar$ ($\rsun$)\dotfill      &  \hatcurISOrlong      & \hatcurisoshort+\hatcurlumind+SME         \\
~~~~$\loggstar$ (cgs)\dotfill       &  \hatcurISOlogg       & \hatcurisoshort+\hatcurlumind+SME         \\
~~~~$\lstar$ ($\lsun$)\dotfill      &  \hatcurISOlum        & \hatcurisoshort+\hatcurlumind+SME         \\
~~~~$M_V$ (mag)\dotfill             &  \hatcurISOmv         & \hatcurisoshort+\hatcurlumind+SME         \\
~~~~$M_K$ (mag,\hatcurjhkfilset)\dotfill &  \hatcurISOMK    & \hatcurisoshort+\hatcurlumind+SME         \\
~~~~Age (Gyr)\dotfill               &  \hatcurISOage        & \hatcurisoshort+\hatcurlumind+SME         \\
~~~~Distance (pc)\dotfill           &  \hatcurXdist         & \hatcurisoshort+\hatcurlumind+SME\\
[-1.5ex]
\enddata
\tablenotetext{a}{
	SME = ``Spectroscopy Made Easy'' package for the analysis of
	high-resolution spectra \citep{valenti:1996}.  These parameters
	rely primarily on SME, but have a small dependence also on the
	iterative analysis incorporating the isochrone search and global
	modeling of the data, as described in the text.
}
\tablenotetext{b}{
	\hatcurisoshort+\hatcurlumind+SME = Based on the \hatcurisoshort\
    isochrones \citep{\hatcurisocite}, \hatcurlumind\ as a luminosity
    indicator, and the SME results.
}
\end{deluxetable}

\subsection{Excluding blend scenarios}
\label{sec:blend}

Our initial spectroscopic analyses discussed in \refsecl{recspec} and
\refsecl{hispec} rule out the most obvious astrophysical false
positive scenarios.  However, more subtle phenomena such as blends
(contamination by an unresolved eclipsing binary, whether in the
background or associated with the target) can still mimic both the
photometric and spectroscopic signatures we see.  In the following
sections we consider and rule out the possibility that such scenarios
may have caused the observed photometric and spectroscopic features.

\subsubsection{Spectral line-bisector analysis}
\label{sec:bisec}

Following \cite{queloz:2001,torres:2007}, we explored the possibility
that the measured radial velocities are not real, but are instead
caused by distortions in the spectral line profiles due to
contamination from a nearby unresolved eclipsing binary.  A bisector
analysis based on the Keck spectra was done as described in \S 5 of
\cite{bakos:2007a}. While the bisector spans show no evidence for
variation in phase with the orbital period, the scatter on these
values ($13.3$\,\ms\ RMS) is large relative to the RV semi-amplitude
($\sim 8$\,\ms), and thus the lack of variation does not provide a
strong constraint on possible blend scenarios. We note that some of
the scatter in the bisector spans may be due to contamination from the
sky background (predominately moonlight)--correcting the bisector
spans for sky contamination as discussed by \citet{hartman:2010}
reduces the RMS to $\sim 10.0$\,\ms, however the precision is still
insufficient to rule out blend scenarios.

\subsubsection{Contamination from a background eclipsing binary}
\label{sec:bgeb}
\begin{figure*}[!ht]
\plotone{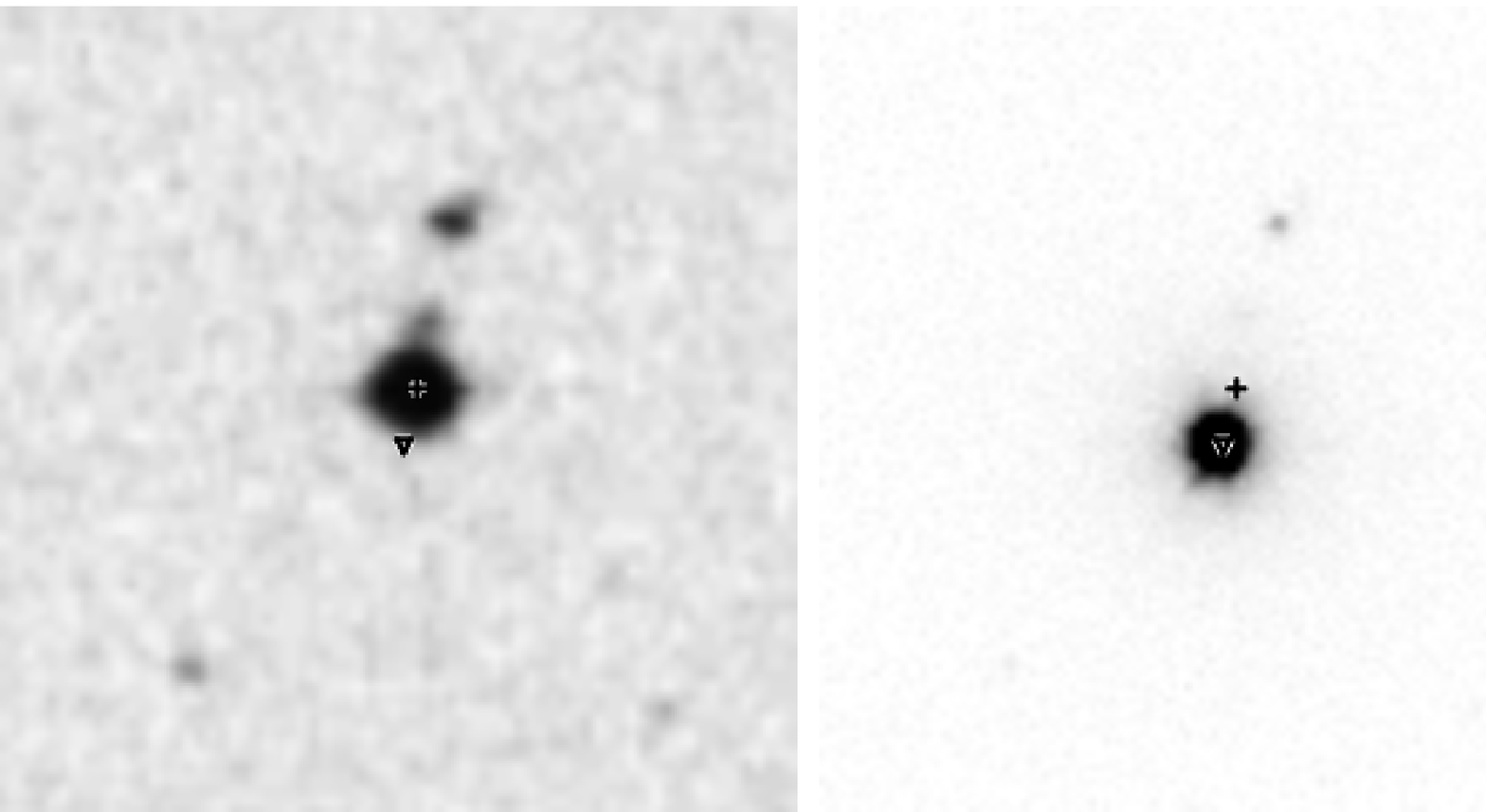}
\caption{
Images of a $2\arcmin \times 2\arcmin$ field containing \hatcur\ from
the POSS-I Red survey (left), and from our \flwof\ \band{i} follow-up
observations (right, see \refsecl{phot}). North is up and east is to
the left in both images. The dates of the exposures are 1950 April 21
and 2010 May 08, respectively. The cross marks the position of
\hatcur\ in 1950 and the triangle marks the position in 2010. Between
these two dates \hatcur\ has moved $\sim 8\farcs9$ to the
southeast. From the POSS-I image, we can rule out the presence of
stars brighter than $R \sim 19$ at the current position of \hatcur.
\label{fig:ppm}}
\end{figure*}

Following our earlier work \citep{bakos:2010,hartman:2009} we make use
of the high proper motion of \hatcur\ to rule out the possibility that
the observed transits and RV variation may be due to a background
eclipsing binary that is aligned, by chance, with the foreground
\hatcurISOspec\ dwarf \hatcur. To reproduce the observed $\sim 0.6\%$
deep transit, the background object cannot be more than 5.6 mag
fainter than \hatcur\ (objects fainter than this would contribute less
than 0.6\% of the total combined light and so could not cause the
transit even if they were to be completely eclipsed by an object that
emits no light). Because \hatcur\ has a high proper motion
\citep[\hatcurCCpm\,\masy;][]{roeser:2010} it is possible to use the
Palomar Observatory Sky Survey plates from 1950 (POSS-I, red and blue
plates) to view the sky at the current position of \hatcur. Between
1950 and the follow-up observations in 2010, \hatcur\ has moved $\sim
8\farcs9$. \reffigl{ppm} shows an image stamp from the POSS-I plate
compared with a recent observation from the \flwof. We can rule out a
background object down to $R \sim 19$\,mag within $\sim 3\arcsec$ of
the current position of \hatcur. Any background object must be $\ga
7.5$ mag fainter than \hatcur\ and thus could not be responsible for
the observed transit.

\subsubsection{Detailed blend modeling of a hierarchical triple}
\label{sec:hitrip}

\begin{figure*}[!ht]
\plotone{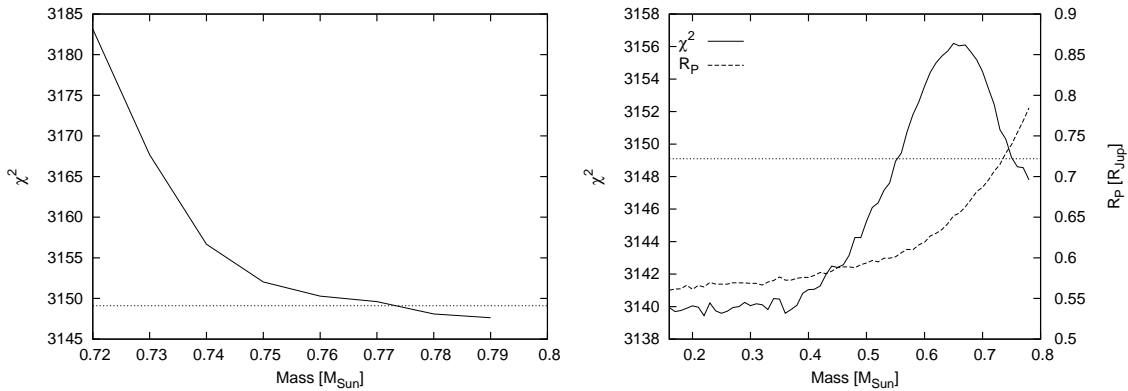}
\caption{
  Left: $\chi^{2}$ for the scenario of a binary star with a planet
  orbiting the fainter star (case 3 in \refsecl{hitrip}) vs. the mass
  of the planet host. Points above the dotted horizontal line are
  rejected at the $3\sigma$ confidence level. The mass of the
  brightest star in the system is fixed to $0.788$\,\msun\ from the
  spectroscopically determined effective temperature, surface gravity
  and metallicity. Right: $\chi^{2}$ for the scenario of a binary star
  with a planet orbiting the brighter star (case 4 in
  \refsecl{hitrip}) vs. the mass of the fainter star. We also show the
  radius of the planet (right axis). Points on the $\chi^{2}$ curve
  above the dotted horizontal line are rejected at the $3\sigma$
  confidence level.
\label{fig:chi2blend}}
\end{figure*}

Following \citet{bakos:2010}, \citet{hartman:2009}, and
\citet{torres:2005} we attempt to model the observations as a
hierarchical triple system. We consider 4 possibilities:
\begin{enumerate}
\item One star orbited by a planet,
\item Three stars, 2 fainter stars are eclipsing,
\item Two stars, 1 planet, planet orbits the fainter star,
\item Two stars, 1 planet, planet orbits the brighter star.
\end{enumerate}
Here case 1 is the fiducial model to which we compare the various
blend models. We model the observed follow-up and HATNet light curves
(including only points that are within one transit duration of the
primary transit or secondary eclipse assuming zero eccentricity)
together with the 2MASS and TASS photometry. In each case we fix the
mass of the brightest star to $0.788$\,\msun; this ensures that we
reproduce the effective temperature, metallicity, and surface gravity
determined from the SME analysis when using the Padova isochrones (see
below). We have also attempted to perform the fits described below
allowing the mass of the brightest star to vary. We find that in this
case the mass of the brightest star is still constrained to be close
to $0.788$\,\msun\ by the broad-band photometry, even if the
spectroscopic parameters are not included. We therefore conclude that
fixing the mass of the brightest star is justified. In all cases we
vary the distance to the system and two parameters allowing for
dilution in the two HATNet light curves, and we include simultaneous
EPD and TFA in fitting the light curves (see \refsecl{globmod}). In
each case we draw the stellar radii and magnitudes from the 13.0\,Gyr
Padova isochrone \citep{girardi:2000}, extended below
0.15\,\msun\ with the \citet{baraffe:1998} isochrones. We use these
rather than the YY isochrones for this analysis because of the need to
allow for stars with $M < 0.4$\,\msun, which is the lower limit
available for the YY models. We use the JKTEBOP program
\citep{southworth:2004a,southworth:2004b} which is based on the
Eclipsing Binary Orbit Program
\citep[EBOP;][]{popper:1981,etzel:1981,nelson:1972} to generate the
model light curves. We optimize the free parameters using the Downhill
Simplex Algorithm together with classical linear least squares for the
EPD and TFA parameters. We rescale the errors for each light curve
such that $\chi^{2}$ per degree of freedom is 1.0 for the out of
transit portion of the light curve. Note that this is done prior to
applying the EPD/TFA corrections, as a result $\chi^{2}$ per degree of
freedom is less than 1.0 for each of the best-fit models discussed
below. If the rescaling is not performed, the difference in $\chi^{2}$
between the best-fit models is even more significant than what is
given below, and the blend-models may be rejected with higher
confidence.

{\em Case 1: 1 star, 1 planet}: In addition to the parameters
mentioned above, in this case we vary the radius of the planet and the
impact parameter of the transit. The best-fit model has $\chi_{\rm
  Case1}^{2} = 3140.1$ for 3567 degrees of freedom. The parameters
that we obtain for the planet are comparable to those obtained from
the global modelling described in \refsecl{globmod}.

{\em Case 2: 3 stars}: For case 2 we vary the masses of the eclipsing
components, and the impact parameter of the eclipse. We find no model
of three stars which reproduces the observations. The transit depth
and duration cannot be fit when the three stars are constrained to
fall on the same isochrone, and the brightest star has $M =
0.788$\,\msun. The best-fit case 2 model consists of equal masses for
the brightest two stars, and $0.08$\,\msun\ (the lowest stellar mass
in the \citealt{baraffe:1998} isochrones) for the transiting
star. Such a model is inconsistent with our spectroscopic observations
(it would have been easily identified as a double-lined binary at one
of the quadrature phases), and as we will show, can be rejected from
the light curves alone. The best-fit case 2 model yields $\chi_{\rm
  Case2}^{2} = 3310.2$ for 3566 degrees of freedom and produces model
transits that are too deep compared to the observed transit. The case
1 model achieves a lower $\chi^{2}$ with fewer parameters than the
case 2 model, so the case 1 model is preferred over the case 2
model. Assuming that the errors are uncorrelated and follow a Gaussian
distribution, the case 2 model can be rejected in favor of the case 1
model at the $> 116\sigma$ confidence level. Alternatively, one might
suppose that any apparent correlations in the residuals of the
best-fit case 2 model are not due to errors in the model but instead
are due to uncorrected systematic errors in the measurements; large
systematic errors in the measurements would increase the probability
that case 1 might give a better fit to the data, by chance, than case
2. To establish the statistical signficance with which we may reject
case 2 while allowing for possible systematic errors in the
measurements, we conduct 1000 Monte Carlo simulations in which we
assume the best-fit case 2 model scenario is correct, shuffle the
residuals from this fit in a manner that preserves the correlations
(this is done by taking the Fourier Transform of the residuals,
randomly changing phases of the Transform while preserving the
amplitudes, and transforming back to the time domain), and fit both
the case 2 and case 1 models to the simulated data. The median value
of $\chi^{2}_{\rm Case2} - \chi^{2}_{\rm Case1}$ is $-17.6$ with a
standard deviation of $26.3$. None of the 1000 trials have
$\chi^{2}_{\rm Case2} - \chi^{2}_{\rm Case1} > 170.1$, the measured
value. Based on this analysis we reject case 2, i.e.~the hierarchical
triple star system scenario, at the $7\sigma$ level.

{\em Case 3: 2 stars, planet orbits the fainter star}: In this
scenario \hatcurb{} is a transiting planet, but it would have a radius
that is larger than what we infer (it may be a Saturn- or Jupiter-size
planet rather than a Neptune-size planet). For this case we vary the
mass of the faint planet-hosting star, the radius of the planet, and
the impact parameter of the transit. We assume the mass of the planet
is negligible relative to the mass of its faint host
star. \reffigl{chi2blend} shows $\chi^{2}$ as a function of the mass
of the planet-hosting star for this scenario. The best-fit case 3
solution has $\chi^{2} = 3147.6$ for 3566 degrees of freedom, and
corresponds to a system where the two stars are of equal mass and the
planet has a radius of $0.8$\,\rjup. As the mass of the planet host is
decreased the value of $\chi^{2}$ increases. Repeating the procedure
outlined above to establish the statistical significance at which we
may reject case 3 we find that the $3\sigma$ limit on $\chi^{2}_{\rm
  Case3} - \chi^{2}_{\rm Case1}$ is $9.0$, which results in a
$3\sigma$ lower-limit of $0.77$\,\msun\ on the mass of the planet
hosting star. We may thereby place a $3\sigma$ lower limit of $0.74$
on the $V$-band luminosity ratio of the two stars. A second set of
lines with a luminosity ratio of $> 0.74$ would have easily been
detectable in both the Keck and TRES spectra unless the stars had very
similar $\gamma$ velocities (the spectral lines are quite narrow with
$\vsini=\hatcurSMEivsin\,\kms$). The poor fit for this blend model
relative to the fiducial model together with the tight constraints on
the relative $\gamma$ velocities and luminosity ratios of the stars in
the blend models that may yet fit the data leads us to reject this
blend scenario in favor of the simpler model of a single star hosting
a transiting planet.

{\em Case 4: 2 stars, planet orbits the brighter star}: As in the
previous case, in this scenario \hatcurb{} is a transiting planet, but
the dilution from the blending star means that the true radius is
larger than what we infer in Case 1. In this case we vary the radius
of the planet, the mass of the faint star, and the impact parameter of
the transit. Again we assume the mass of the planet is negligible
relative to the mass of its bright host star. \reffigl{chi2blend}
shows $\chi^{2}$ as a function of the mass of the faint star. The
smallest value of $\chi^{2}$ is achieved when the faint star
contributes negligible light to the system, which is effectively
equivalent to the fiducial scenario represented by Case 1. Two effects
cause $\chi^{2}$ to increase with stellar mass. First, the shape of
the transit is subtly changed in a manner that gives a poorer fit to
the observations. Second, when the mass of the faint star is less than
that of the transit host the model broad-band photometry for the
blended system is redder than for the single star scenario, and is
inconsistent with the observed photometry. This gives rise to the peak
in $\chi^2$ at $M \sim 0.65$\,\msun. The case 4 model where the faint
star has $M \ga 0.77$\,\msun\ can be rejected as in Case 3. For lower
masses, we place a $3\sigma$ upper limit of 0.55\,\msun\ on the mass
of the faint star, yielding a $3\sigma$ upper limit on the luminosity
ratio of $0.1$. We conclude that at most the planet radius $R_{p}$ may
be 8\% larger than what we find in \refsecl{globmod} if there is an
undetected faint secondary star in the system.


\subsection{Global modeling of the data}
\label{sec:globmod}

Here we summarize the procedure that we followed to model the HATNet
photometry, the follow-up photometry, and the radial velocities
simultaneously. This procedure is described in greater detail in
\citet{bakos:2010}. The follow-up \lcs\ were modeled using analytic
formulae based on \citet{mandel:2002}, with quadratic limb darkening
coefficients for the Sloan \band{i} interpolated from the tables by
\citet{claret:2004} for the spectroscopic parameters of the star as
determined from the SME analysis (\refsecl{stelparam}). We modeled the
HATNet data using an approximation to the \citet{mandel:2002} formulae
as described in \citet{bakos:2010}. The RVs were fitted with an
eccentric Keplerian model.

Our physical model consisted of 8 main parameters, including: the time
of the first transit center observed with HATNet (taken to be event
$-74$), $T_{c,-74}$, and that of the last transit center observed with
the \flwof\ telescope, $T_{c,+40}$, the normalized planetary radius
$p\equiv \rpl/\rstar$, the square of the impact parameter $b^2$, the
reciprocal of the half duration of the transit $\zrstar$ as given in
\citet{bakos:2010}, the RV semiamplitude $K$, and the Lagrangian
elements $k \equiv e \cos \omega$ and $h \equiv e \sin \omega$, where
$\omega$ is the longitude of periastron. Five additional parameters
were included that have to do with the instrumental
configuration. These are the HATNet blend factors $B_{\rm inst,376}$,
and $B_{\rm inst,377}$, which account for possible dilution of the
transit in the HATNet \lcs\ from background stars due to the broad PSF
($\sim 24\arcsec$ FWHM), the out-of-transit magnitudes for each HATNet
field, $M_{\rm 0,HATNet,376}$ and $M_{\rm 0,HATNet,377}$, and the
relative zero-point $\gamma_{\rm rel}$ of the Keck RVs. The physical
model was extended with an instrumental model for the follow-up light
curves that describes brightness variations caused by systematic
errors in the measurements. We adopted a ``local'' EPD- and ``global''
TFA-model \citep{bakos:2010}, using 20 template stars
for the TFA procedure and six EPD parameters for each follow-up light
curve. In summary, the total number of fitted parameters was 13
(physical model with 5 configuration-related parameters) + 30 (local
EPD) + 20 (global TFA) = 63, i.e., much smaller than the number of
data points (1450, counting only RV measurements and follow-up
photometry measurements).

As described in \citet{bakos:2010}, we use a combination of the
downhill simplex method \citep[AMOEBA; see][]{press:1992}, the
classical linear least squares algorithm, and the Markov Chain
Monte-Carlo method \citep[MCMC, see][]{ford:2006} to obtain a best-fit
model together with {\em a posteriori} distributions for the fitted
parameters. These distributions were then used to obtain {\em a
  posteriori} distributions for other quantities of interest, such as
$\rhostar$. As described in \refsecl{stelparam}, $\rhostar$ was used
together with stellar evolution models to obtain {\em a posteriori}
distributions for stellar parameters, such as $\mstar$ and $\rstar$,
which are needed to determine $\mpl$ and $\rpl$.

The resulting parameters pertaining to the light curves and velocity
curves, together with derived physical parameters of the planet, are
listed under the ``Adopted Value'' column heading of
\reftabl{planetparam}. Included in this table is the RV
``jitter''. This is a component of assumed astrophysical noise
intrinsic to the star, possibly with a contribution from instrumental
errors as well, that we added in quadrature to the internal errors for
the RVs in order to achieve $\chi^{2}/{\rm dof} = 1$ from the RV data
for the global fit.  Auxiliary parameters not listed in the table are:
$T_{\mathrm{c},-74}=\hatcurLCTA$~(BJD),
$T_{\mathrm{c},+40}=\hatcurLCTB$~(BJD), the blending factors
$B_{\rm instr,376}=\hatcurLCiblendA$, and $B_{\rm instr,377}=\hatcurLCiblendB$
for the HATNet field 376 and 377 light curves, respectively, and
$\gamma_{\rm rel}=\hatcurRVgamma$\,\ms.
The latter quantity represents an arbitrary offset for the Keck RVs,
but does not correspond to the true center-of-mass velocity of
the system, which was listed earlier as $\gamma_{\rm RV}$ in 
\reftabl{stellar}.

We find a mass for the planet of
$\mpl=\hatcurPPmlong\,\mjup$ and a radius of
$\rpl=\hatcurPPrlong\,\rjup$, leading to a mean density
$\rho_p=\hatcurPPrho$\,\gcmc. 
We also find that the eccentricity of the orbit may be different from
zero: $e = \hatcurRVeccen$, $\omega = \hatcurRVomega\arcdeg$. However,
as we show in \refsecl{eccen}, this is at best significant at only the
88\% confidence level.

We also carried out the analysis described above with the eccentricity
fixed to zero. The resulting parameters are given in
\reftabl{planetparam} under the column heading
``$\{\zeta/\rstar,b^2,p\}, e \equiv 0$''. The results are discussed
further in \refsecl{eccen}.

Finally, we conducted an independent model of the system based on
\citet{kipping:2010b}. The primary differences between this model and
the adopted model are differences in the choice of parameters to vary
in the fit: we use $\Upsilon / \rstar$ as defined in
\citet{kipping:2010b} rather than $\zeta / \rstar$, $b$ rather than
$b^{2}$, and $p^2$ rather than $p$. We also allowed for a linear drift
in the radial velocities $\dot{\gamma}$, and a time shift $t_{\rm
  troj}$ in the radial velocities due to possible additional bodies in
the system on Trojan orbits with \hatcurb. We chose to include both
$t_{\rm troj}$ and $\dot{\gamma}$ rather than fixing them to zero as
the value of the Bayseian Information Criterion \citep[BIC;
  e.g.][]{kipping:2010c} was lower for the best-fit model including
these parameters, than for models where one or both of these
parameters were fixed to zero. The resulting parameters are given in
\reftabl{planetparam} under the column heading
``$\{\Upsilon/\rstar,b,p^2\}$.''  The parameter values from this model
are consistent with those from the adopted model, which gives
confidence that our results are robust to changes in the choice of
fitting parameters.

\ifthenelse{\boolean{emulateapj}}{
    \begin{deluxetable*}{lrrr}
}{
    \begin{deluxetable}{lrrr}
}
\tabletypesize{\scriptsize}
\tablecaption{Orbital and planetary parameters\label{tab:planetparam}}
\tablehead{
	\multicolumn{1}{c}{~~~~~~~~~~~~~~~Parameter~~~~~~~~~~~~~~~} &
	\multicolumn{1}{c}{{\bf Adopted Value}} &
        \multicolumn{1}{c}{Value} &
        \multicolumn{1}{c}{Value} \\
        \multicolumn{1}{c}{} &
        \multicolumn{1}{c}{$\{\zeta/\rstar,b^2,p\}$} &
        \multicolumn{1}{c}{$\{\zeta/\rstar,b^2,p\}, e \equiv 0$} &
        \multicolumn{1}{c}{$\{\Upsilon/\rstar,b,p^2\}$} 
}
\startdata
\noalign{\vskip -3pt}
\sidehead{\Lc{} parameters}
~~~$P$ (days)             \dotfill    & $\hatcurLCP$        & $\hatcurnoeccenLCP$  & $4.234508_{-0.000022}^{+0.000021}$             \\
~~~$T_c$ (${\rm BJD}$)    
      \tablenotemark{a}   \dotfill    & $\hatcurLCT$        & $\hatcurnoeccenLCT$  & $2455304.65120_{-0.00049}^{+0.00048}$            \\
~~~$T_{14}$ (days)
      \tablenotemark{a}   \dotfill    & $\hatcurLCdur$        & $\hatcurnoeccenLCdur$ & $0.1023_{-0.0011}^{+0.0012}$          \\
~~~$T_{12} = T_{34}$ (days)
      \tablenotemark{a}   \dotfill    & $\hatcurLCingdur$        & $\hatcurnoeccenLCingdur$  & $0.00724_{-0.00027}^{+0.00081}$       \\
~~~$\arstar$              \dotfill    & $\hatcurPPar$        & $\hatcurnoeccenPPar$ & $13.28_{-0.76}^{+0.70}$            \\
~~~$\zrstar$              \dotfill    & $\hatcurLCzeta$        & $\hatcurnoeccenLCzeta$ & $21.10_{-0.21}^{+0.20}$          \\
~~~$\rpl/\rstar$          \dotfill    & $\hatcurLCrprstar$        & $\hatcurnoeccenLCrprstar$  & $0.07341_{-0.00093}^{+0.00104}$      \\
~~~$b^2$                  \dotfill    & $\hatcurLCbsq$        & $\hatcurnoeccenLCbsq$  & $0.035_{-0.032}^{-0.089}$          \\
~~~$b \equiv a \cos i/\rstar$
                          \dotfill    & $\hatcurLCimp$        & $\hatcurnoeccenLCimp$  & $0.00_{-0.00}^{+0.26}$          \\
~~~$i$ (deg)              \dotfill    & $\hatcurPPi$        & $\hatcurnoeccenPPi$   & $89.14_{-0.72}^{+0.59}$           \\

\sidehead{Limb-darkening coefficients \tablenotemark{b}}
~~~$a_i$ (linear term)    \dotfill    & $\hatcurLBii$        & $\hatcurnoeccenLBii$  & $0.3862$           \\
~~~$b_i$ (quadratic term) \dotfill    & $\hatcurLBiii$        & $\hatcurnoeccenLBiii$  & $0.2576$          \\

\sidehead{RV parameters}
~~~$K$ (\ms)              \dotfill    & $\hatcurRVK$        & $\hatcurnoeccenRVK$   & $7.6_{-1.2}^{+1.2}$           \\
~~~$k_{\rm RV}$\tablenotemark{c} 
                          \dotfill    & $\hatcurRVk$        & $\hatcurnoeccenRVk$  & $0.09_{-0.11}^{+0.12}$            \\
~~~$h_{\rm RV}$\tablenotemark{c}
                          \dotfill    & $\hatcurRVh$        & $\hatcurnoeccenRVh$  & $0.028_{-0.060}^{+0.063}$           \\
~~~$e$                    \dotfill    & $\hatcurRVeccen$        & $\hatcurnoeccenRVeccen$  & $0.127_{-0.068}^{+0.094}$        \\
~~~$\omega$ (deg)         \dotfill    & $\hatcurRVomega$        & $\hatcurnoeccenRVomega$  & $74_{-59}^{+266}$        \\
~~~$\dot{\gamma}$ (m\,s$^{-1}$\,d$^{-1}$) & $0$ & $0$ & $-0.028_{-0.013}^{+0.014}$ \\
~~~$t_{\rm troj}$ (d)\tablenotemark{d} & $0$ & $0$ & $0.01_{-0.23}^{+0.24}$ \\
~~~RV jitter (\ms)        \dotfill    & \hatcurRVjitter         & \hatcurnoeccenRVjitter & $\ldots$  \\
~~~RV fit RMS (\ms)       \dotfill    & \hatcurRVfitrms         & \hatcurnoeccenRVfitrms & $\ldots$  \\

\sidehead{Secondary eclipse parameters}
~~~$T_s$ (BJD)            \dotfill    & $\hatcurXsecondary$        & $\hatcurnoeccenXsecondary$ & $2455307.01_{-0.29}^{+0.31}$      \\
~~~$T_{s,14}$             \dotfill    & $\hatcurXsecdur$        & $\hatcurnoeccenXsecdur$  & $0.108_{-0.011}^{+0.014}$        \\
~~~$T_{s,12}$             \dotfill    & $\hatcurXsecingdur$        & $\hatcurnoeccenXsecingdur$ & $0.00782_{-0.00083}^{+0.00104}$     \\

\sidehead{Planetary parameters}
~~~$\mpl$ ($\mjup$)       \dotfill    & $\hatcurPPmlong$        & $\hatcurnoeccenPPmlong$ & $0.0522_{-0.0083}^{+0.0084}$        \\
~~~$\rpl$ ($\rjup$)       \dotfill    & $\hatcurPPrlong$        & $\hatcurnoeccenPPrlong$ & $0.553_{-0.031}^{+0.037}$         \\
~~~$C(\mpl,\rpl)$
    \tablenotemark{e}     \dotfill    & $\hatcurPPmrcorr$        & $\hatcurnoeccenPPmrcorr$  & $0.059$       \\
~~~$\rhopl$ (\gcmc)       \dotfill    & $\hatcurPPrho$        & $\hatcurnoeccenPPrho$   & $0.378_{-0.084}^{+0.099}$         \\
~~~$\log g_p$ (cgs)       \dotfill    & $\hatcurPPlogg$        & $\hatcurnoeccenPPlogg$ & $2.621_{-0.092}^{+0.084}$          \\
~~~$a$ (AU)               \dotfill    & $\hatcurPParel$        & $\hatcurnoeccenPParel$ & $0.04780_{-0.00061}^{+0.00064}$          \\
~~~$T_{\rm eq}$ (K)       \dotfill    & $\hatcurPPteff$        & $\hatcurnoeccenPPteff$ & $991_{-36}^{+42}$          \\
~~~$\Theta$\tablenotemark{f}\dotfill  & $\hatcurPPtheta$        & $\hatcurnoeccenPPtheta$ & $0.0110_{-0.0019}^{+0.0020}$         \\
~~~$F_{peri}$ ($10^{\hatcurPPfluxperidim}$\ergscmsq) \tablenotemark{g}
                          \dotfill    & $\hatcurPPfluxperi$        & $\hatcurnoeccenPPfluxperi$ & $2.87_{-0.63}^{+1.03}$      \\
~~~$F_{ap}$  ($10^{\hatcurPPfluxapdim}$\ergscmsq) \tablenotemark{g} 
                          \dotfill    & $\hatcurPPfluxap$        & $\hatcurnoeccenPPfluxap$ & $1.72_{-0.31}^{+0.28}$        \\
~~~$\langle F \rangle$ ($10^{\hatcurPPfluxavgdim}$\ergscmsq) 
\tablenotemark{g}         \dotfill    & $\hatcurPPfluxavg$        & $\hatcurnoeccenPPfluxavg$ & $2.18_{-0.30}^{+0.40}$       \\
[-1.5ex]
\enddata
\tablenotetext{a}{
    \ensuremath{T_c}: Reference epoch of mid transit that minimizes the
    correlation with the orbital period.  It corresponds to $N_{tr} =
    +31$. BJD is calculated from UTC.
	\ensuremath{T_{14}}: total transit duration, time between first to
	last contact;
	\ensuremath{T_{12}=T_{34}}: ingress/egress time, time between first
	and second, or third and fourth contact.
}
\tablenotetext{b}{
	Values for a quadratic law, adopted from the tabulations by
    \cite{claret:2004} according to the spectroscopic (SME) parameters
    listed in \reftabl{stellar}.
}
\tablenotetext{c}{
    Lagrangian orbital parameters derived from the global modeling, and
    primarily determined by the RV data.
}
\tablenotetext{d}{
        Time-offset in the radial velocities due to companion planets in Trojan orbits.
}
\tablenotetext{e}{
	Correlation coefficient between the planetary mass \mpl\ and radius
	\rpl.
}
\tablenotetext{f}{
	The Safronov number is given by $\Theta = \frac{1}{2}(V_{\rm
	esc}/V_{\rm orb})^2 = (a/\rpl)(\mpl / \mstar )$
	\citep[see][]{hansen:2007}.
}
\tablenotetext{g}{
	Incoming flux per unit surface area, averaged over the orbit.
}
\ifthenelse{\boolean{emulateapj}}{
    \end{deluxetable*}
}{
    \end{deluxetable}
}



\section{Discussion}
\label{sec:discussion}

We have presented the discovery of \hatcurb, a transiting Neptune-mass
planet. Below we discuss the physical properties of this planet, and
compare them to the properties of similar planets; we comment on the
possibility that the planet has undergone significant evaporation, on
the significance of its orbital eccentricity, and on the possible
presence of additional bodies in the system; and we discuss the
prospects for detailed follow-up studies.

\subsection{Physical Properties of \hatcurb}
\begin{figure*}[!ht]
\plotone{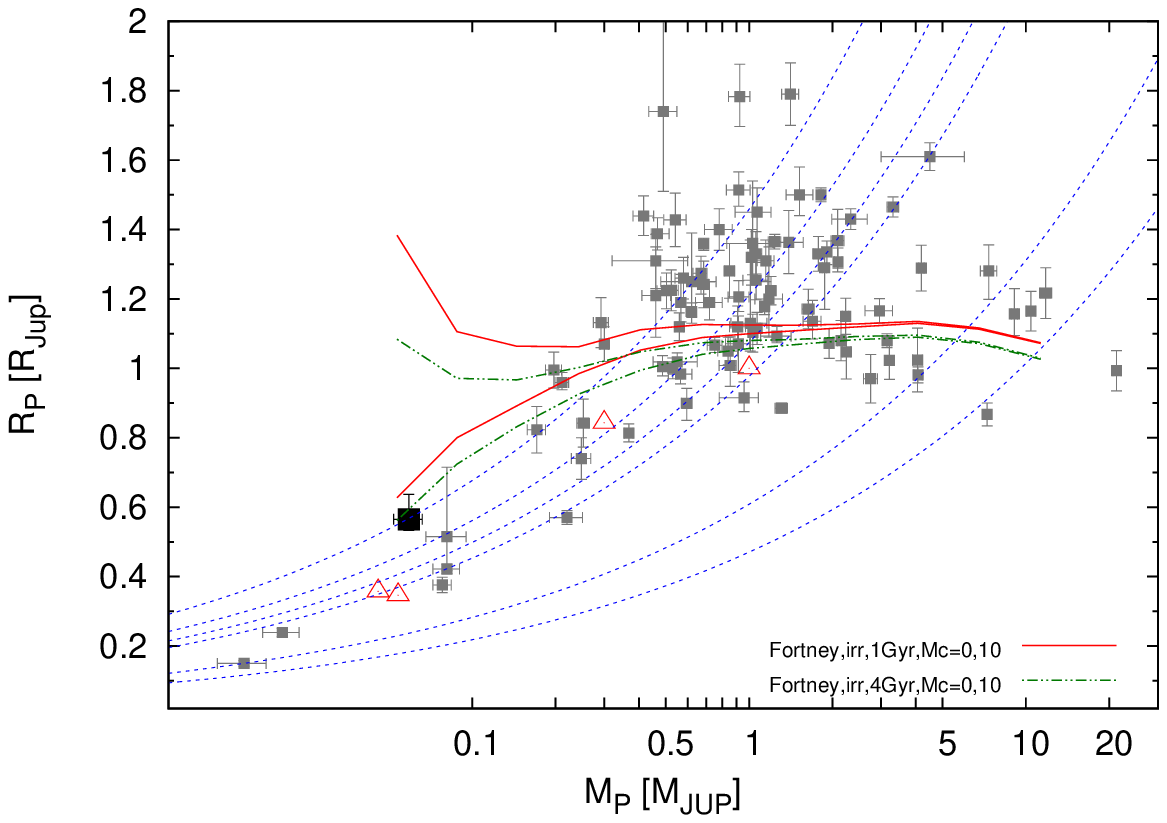}
\plotone{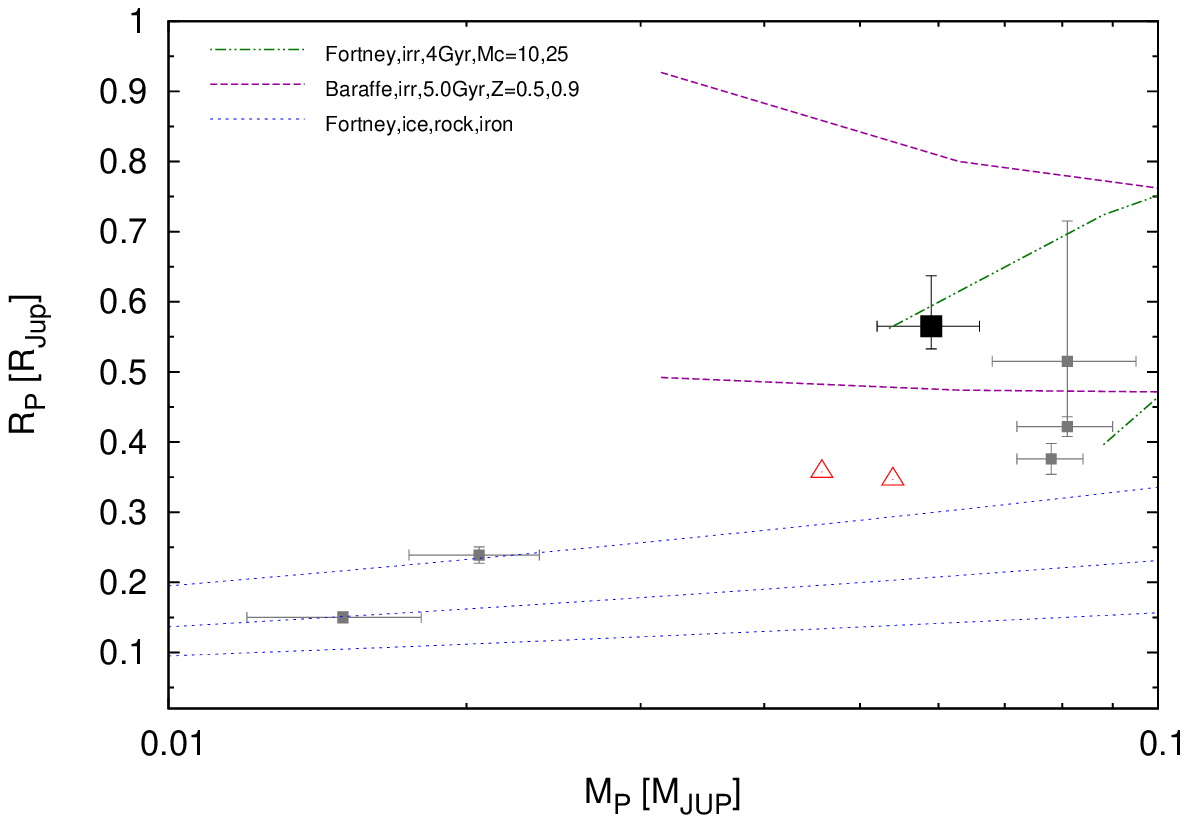}
\caption{
  Top: Mass--radius diagram of known TEPs (small filled squares).
  \hatcurb\ is shown as a large filled square.  Overlaid are
  \citet{fortney:2007} theoretical planetary mass--radius curves
  interpolated to the solar equivalent semi-major axis of \hatcurb{}
  for ages of 1.0\,Gyr (upper, solid lines) and 4\,Gyr (lower
  dashed-dotted lines) and core masses of 0 and 10\,\mearth (upper and
  lower lines respectively), as well as isodensity lines for 0.4, 0.7,
  1.0, 1.33, 5.5 and 11.9\,\gcmc (dashed lines).  Solar system planets
  are shown with open triangles. Bottom: Mass--radius diagram for
  planets with $0.01\,\mjup < M_{p} < 0.1\,\mjup$. References for the
  low-mass planet parameters are given in \refsecl{introduction} of
  the text. We adopt the ``a.e'' model from \citet{kipping:2010} for
  Kepler-4b. Overlaid are the interpolated \citet{fortney:2007}
  theoretical relations for 4\,Gyr and core masses of 10 and
  25\,\mearth, \citet{fortney:2007} theoretical curves for pure ice,
  rock and iron composition planets (upper, middle, and lower dotted
  lines), and the \citet{baraffe:2008} theoretical irradiated curves
  for heavy element mass fractions of $Z = 0.5$ and $0.9$ (upper and
  lower dashed lines respectively).
\label{fig:exomr}}
\end{figure*}

\reffigl{exomr} compares \hatcurb\ to the other known TEPs on a
mass-radius diagram. With a density of \hatcurPPrho\,\gcmc,
\hatcurb\ is significantly less dense than the four other Neptune-size
planets with well measured masses and radii (Uranus, Neptune, GJ~436b,
HAT-P-11b). For Kepler-4b, \citet{kipping:2010} find a large
uncertainty on the radius which results from significant uncertainties
on the eccentricity and the transit impact parameter. Kepler-4b may be
comparable in size to GJ~436b and HAT-P-11b, or it could be even less
dense than \hatcurb.

From the theoretical models of \citet{fortney:2007}, \hatcurb\ has a
radius that is well above the maximum radius of $0.3$\,\rjup\ for a
\hatcurPPmshort\,\mjup\ planet lacking a hydrogen-helium envelope
(i.e.~a planet with a 100\% water-ice composition). The best-fit mass
and radius for \hatcurb\ falls just below the 4\,Gyr model with a
10\,\mearth\ rocky core and 8\,\mearth\ gas envelope, implying that a
4\,Gyr model with a slightly higher core mass would provide a better
match to the mass and radius. We note that the isochrone-based age
(\hatcurISOage\,Gyr) and the activity-based age (\hatcurRHKgyroage\,Gyr) for the
\hatcur\ system are somewhat older than 4\,Gyr, so the inferred
core-mass would therefore be somewhat smaller.

We also compare \hatcurb\ to the theoretical models of
\citet{baraffe:2008} which predict more significant inflation due to
irradiation for low mass planets than do the \citet{fortney:2007}
models. In this case the radius of \hatcurb\ is intermediate between
the $Z = 0.5$ and $Z = 0.9$ heavy-element enrichment models.

\subsection{Evaporation}

Observations of the transiting hot Jupiters HD~209458b and HD~189733b
in the \ion{H}{1} Lyman-$\alpha$ line have indicated that both planets
are evaporating at a rate of up to $\sim 10^{10}$\,g\,s$^{-1}$
\citep[e.g,][]{vidalmadjar:2003,lecavelierdesetangs:2010}. Prompted by
the observations for HD~209458b, several theoretical studies have
indicated that atmospheric evaporation is likely to be important for
close-in planets, particularly those with low surface gravities, such
as hot Neptunes (see for example \citealp{lammer:2003} and the review
by \citealp{yelle:2008}). It has even been suggested that some
close-in Neptune-mass and smaller planets may be the evaporated cores
of planets which initially had masses comparable to Saturn or Jupiter
\citep[e.g.][]{baraffe:2005}. In the case of energy-limited escape,
the evaporative mass-loss is given by (see \citealp{erkaev:2007}, and
\citealp{yelle:2008}; see also \citealp{valencia:2010} and
\citealp{jackson:2010} for applications to CoRoT-7b):
\begin{equation}
\dot{M_{p}} = -\frac{\pi R_{p}^{3}\epsilon F_{\rm XUV}}{G M_{p}K_{\rm tide}}
\label{eqn:mdotevap}
\end{equation}
where $F_{\rm XUV}$ is the incident flux of extreme ultraviolet (XUV)
stellar radiation, $\epsilon$ is the heating efficiency and is
estimated to be $\sim 0.4$ for the case of HD~209458b
\citep{yelle:2008}, and $K_{\rm tide}$ is a factor that accounts for
an enhancement of the evaporation rate in the presence of tides, and
is given by:
\begin{equation}
K_{\rm tide} = 1 - \frac{3}{2\xi} + \frac{1}{2\xi^{3}}
\end{equation}
where $\xi = (M_{p}/(3M_{\star}))^{1/3}a/R_{p}$ is the ratio of the
Roche radius to the planet radius. \citet{ribas:2005} find that for
solar type stars the XUV flux at 1 AU integrated over the wavelength
range $1$\,\AA\ to $1200$\,\AA\ is given by:
\begin{equation}
F_{\rm XUV,1AU} = 29.7\tau^{-1.23}\,{\rm ergs}\,{\rm s}^{-1}\,{\rm cm}^{-2}
\end{equation}
where $\tau$ is the age in Gyr. To our knowledge, a similar study has
not been completed for K dwarfs, however long term X-ray observations
of the 5-6\,Gyr $\alpha$~Cen~AB system reveal that on average the K1
dwarf star $\alpha$~Cen~B has an X-ray luminosity in the
6-60\,\AA\ band that is approximately twice that of the Sun, while the
G2 dwarf $\alpha$~Cen~A has a luminosity that is approximately half
that of the Sun \citep{ayers:2009}. For simplicity we therefore assume
that the total XUV luminosity of \hatcur\ is comparable to that of the
Sun \citep[4.64\,ergs\,s$^{-1}$\,cm$^{-2}$ at 1 AU;][]{ribas:2005},
which is likely correct to within an order of magnitude. Assuming
$\epsilon = 0.4$, we estimate that the expected present-day mass-loss
rate for \hatcurb\ is $\sim 3\times 10^{10}\,{\rm g}\,{\rm s}^{-1} =
0.17\,\mearth\,{\rm Gyr}^{-1}$. To determine the total mass lost by
\hatcurb\ over its lifetime, we integrate equation~\ref{eqn:mdotevap}
assuming an age of $4.5$\,Gyr, $F \propto \tau^{-1.23}$ for $\tau >
0.1$\,Gyr and $F \equiv {\rm constant}$ for $\tau < 0.1$\,Gyr,
neglecting tidal evolution of the orbit, and assuming that the radius
is constant. We find that \hatcurb\ may have lost a significant
fraction its mass ($\sim$ 30\%); the exact value depends strongly on
several poorly constrained parameters including $F_{\rm XUV}$ and its
dependence on age for a K1 dwarf, $\epsilon$, and the age of the
system.

\subsection{Eccentricity}
\label{sec:eccen}

Using the relation given by \citet{adams:2006}, the expected
tidal circularization time-scale for \hatcurb\ is $\sim 1$\,Gyr which
is much less than the age of the system. This time-scale is estimated
assuming a large tidal quality factor of $Q_{P} = 10^6$, and that
there are no additional bodies in the system exciting the
eccentricity. However, because at least two of the three hot Neptunes
have significant eccentricities (GJ~436b has $e = 0.14 \pm 0.01$,
\citealp{demory:2007}; and HAT-P-11b has $e = 0.198 \pm 0.046$,
\citealp{bakos:2010}; the eccentricity for Kepler-4b is poorly
constrained, \citealp{kipping:2010}), we cannot conclude that the
eccentricity must be zero on physical grounds, and therefore do not
adopt a zero-eccentricity model for the parameter
determination. 

As discussed in \refsecl{stelparam} the eccentricity of \hatcurb\ is
poorly constrained by the RV observations, and is instead constrained
by requiring that the star be younger than the age of the universe
(without the age constraint we get $e = 0.24 \pm 0.12$, whereas
including the age constraint gives $e = \hatcurRVeccen$). To establish
the significance of the eccentricity measurement, we also fit a model
with the eccentricity fixed to zero. An F-test
\citep[e.g.][]{lupton:1993} allows us to reject the null hypothesis of
zero eccentricity with only 79\% confidence. Alternatively, the
\citet{lucy:1971} test for the significance of an eccentricity
measurement gives a false alarm probability of $\sim 12$\% for
detecting $e > 0.124$ with an error of $0.060$, or 88\% confidence
that the orbit is eccentric. If the eccentricity is fixed to zero, the
required jitter to achieve $\chi^{2}/N_{\rm dof} = 1$ is 2.4\,\ms,
which is closer to the typical Keck/HIRES jitter of other
chromospherically quiet early K dwarfs than the jitter of
$\hatcurRVjitter$\,\ms\ that is obtained with an eccentric orbit
fit. We therefore are not able to claim a significant eccentricity for
\hatcurb, and instead may only place a $95\%$ confidence upper limit
of $e < 0.22$. For the $\{\Upsilon/\rstar,b,p^2\}$ model discussed at
the end of \refsecl{globmod}, the $95\%$ confidence upper limit is $e
< 0.32$. Further RV observations, or a photometric detection of the
occultation of the planet by its host star, are needed to determine if
the eccentricity is nonzero.

\subsection{Additional Bodies in the System}

The $\{\Upsilon/\rstar,b,p^2\}$ model discussed in
\refsecl{globmod} with parameters given in \reftabl{planetparam}
includes a linear drift in the radial velocities, $\dot{\gamma}$, as a
free parameter. We find $\dot{\gamma} =
-0.028_{-0.013}^{+0.014}$\,m\,s$^{-1}$\,d$^{-1}$. Conducting an odds
ratio test \citep[see][]{kipping:2010c}, we conclude that the drift is
real with 96.6\% confidence, making this a $2.1\sigma$
detection. While the detection is not significant enough for us to be
highly confident that there is at least one additional body in the
system, this suggestive result implies that \hatcur\ warrants
long-term RV monitoring. We also searched for a linear time-shift in
the RVs due to potential Trojans. We do not detect a significant
shift, and may exclude $|t_{\rm troj}| < 0.50$\,d with 95\%
confidence. This translates to an upper limit of $84$\,\mearth\ on the
mass of a Trojan companion, which is greater than the mass of the
planet. With the present data we are thus not able to place a
meaningful limit on the presence of Trojan companions.

\subsection{Suitability for Follow-up}

\hatcur\ has a number of features that make it an attractive target
for potential follow-up studies. At $V = \hatcurCCtassmv$, it is
bright enough that precision spectroscopic and photometric
observations are feasible with moderate integration times. The
equatorial declination of $\delta = \hatcurCCdec$ also means that
\hatcur\ is accessible to both Northern and Southern ground-based
facilities. The exceptionally low jitter will facilitate further RV
observations, which might be used to confirm and refine the
eccentricity determination, to measure the Rossiter-McLaughlin effect
(R-M; discussed in more detail below) and to search for additional
planets in the system.

A detection of the occultation of \hatcurb\ by \hatcur\ with IRAC/{\em
  Spitzer} would provide a strong constraint on $k \equiv e \cos
\omega$, while the duration of the occultation would provide a
constraint on $h \equiv e \sin \omega$. We note that the median value
of the {\em a posteriori} distribution for the time of occultation
that results from our global fit when the eccentricity is allowed to
vary (\refsecl{globmod}) is $6.5$\,h after the expected time of
occultation assuming a circular orbit. The expected depth of the
occultation event is a challenging $0.012\%$ and $0.020\%$ at
3.6\,\micron\ and 4.5\,\micron\ respectively. Scaling from TrES-4, a
somewhat fainter star at these wavelengths, for which
\citet{knutson:2009} measured occultations at 3.6\,\micron\ and
4.5\,\micron\ using IRAC/{\em Spitzer} with precisions of $0.011\%$
and $0.016\%$ respectively, one may hope to achieve a $\sim 1.4\sigma$
and $1.6\sigma$ detection for \hatcurb\ for one event at each
bandpass. One would need to observe 5 and 4 occultations respectively
to achieve a $3\sigma$ detection.

Recent measurements of the R-M effect for TEPs have revealed a
substantial population of planets on orbits that are significantly
misaligned with the spin axes of their host stars
\citep[e.g.][]{triaud:2010}. \citet{winn:2010a} note that misalignment
appears to be more prevalent for planets orbiting stars with $T_{\rm
  eff} > 6250$\,K, and suggest that most close-in planets migrate by
planet-planet or planet-star scattering mechanisms, or by the Kozai
effect, rather than disk migration, and that tidal dissipation in the
convective surfaces of cooler stars realigns the stellar spin axis to
the orbital axis of the close-in massive
planet. \citet{schlaufman:2010} also finds evidence that planets
orbiting stars with $M_{\star} > 1.2$\,\msun\ are more likely to be
misaligned than planets orbiting cooler stars using a method that is
independent of the R-M measurements. One prediction of the
\citet{winn:2010a} hypothesis is that lower mass planets orbiting cool
stars should show a greater degree of misalignment than higher mass
planets due to their reduced tidal influence. The detection of
misalignment for HAT-P-11b \citep{winn:2010b,hirano:2010} is
consistent with this hypothesis. Measuring the R-M effect for
\hatcurb\ would provide an additional test. Using equation (40) from
\citet{winn:2010c}, the expected maximum amplitude of the R-M effect
for \hatcurb\ is $\sim 9$\,\ms, which given the low jitter of \hatcur,
should be detectable at $\ga 10\sigma$.

By measuring the primary transit depth as a function of wavelength it
is possible to obtain a transmission spectrum of an exoplanet's
atmosphere. Such observations have been made for a handful of planets
(e.g.~\citealp{charbonneau:2002}; see also the review by
\citealp{seager:2010}). Following \citet{brown:2001}, the expected
difference in transit depth between two wavelengths is given
approximately by:
\begin{equation}
\Delta\delta = \frac{2R_{p}H}{R_{\star}^2}N_{H}
\end{equation}
where $H = k_{\rm B}T_{\rm eq}/g_{p}\mu$ is the scale height of the
atmosphere, $g_{p}$ is the planet surface gravity, $\mu$ is the mean
molecular weight of the atmosphere, and $N_{H} =
\ln(\sigma_{1}/\sigma_{2})$ where $\sigma_{1}$ and $\sigma_{2}$ are
the opacities per gram of material at wavelengths in a strong atomic
or molecular line and in the nearby continuum respectively. Assuming a
pure H$_2$ atmosphere, $\mu = 3.347 \times 10^{-27}$\,kg, we find for
\hatcurb\ $H = 920$\,km, and $\Delta\delta = 0.0246N_{H}\%$. If
instead we assume that the atmosphere has the same composition as
Neptune \citep[e.g.][]{depater:2001}, we have $\mu = 4.655 \times
10^{-27}$\,kg, $H = 660$\,km, and $\Delta\delta = 0.0177N_{H}\%$. For
comparison, assuming a pure H$_2$ atmosphere, the planet HD~209458b
has $\Delta\delta = 0.0198N_{H}\%$, while GJ~436b has $\Delta\delta =
0.0107N_{H}\%$, HAT-P-11b has $\Delta\delta = 0.0072N_{H}\%$ and
Kepler-4b has $\Delta\delta = 0.0038N_{H}\%$. Due to its low surface
gravity, \hatcurb\ easily has the highest expected transmission
spectrum signal among the known transiting Neptune-mass planets. While
it is relatively faint compared to the well studied planets HD~209458b
and HD~189733b, we note that \citet{sing:2010} used the Gran
Telescopio Canarias (GTC) to detect a $0.058 \pm 0.016\%$ absorption
level at 7582\,\AA\ due to Potassium in the atmosphere of XO-2b, which
orbits a $V = 11.2$ early K star. Scaling from this observation, it
should be possible to detect components in the atmosphere of
\hatcurb\ with $N_{H} \ga 3$ at the $\sim 3\sigma$ level using the
GTC.

\subsection{Summary}

In summary, \hatcurb\ is a low-density Neptune-mass planet. Its
low-density relative to the other known Neptune-mass planets means
that \hatcurb\ likely has a more significant hydrogen-helium gas
envelope than its counterparts. The existence of \hatcurb\ provides
empirical evidence that, like hot Jupiters, hot Neptunes also exhibit
a wide range of densities. Comparing to the \citet{fortney:2007}
models, we find that \hatcurb\ is likely composed of a gas envelope
and a heavy-element core that are approximately equal in mass, while
the \citet{baraffe:2008} models prefer a higher heavy-element
fraction. It is also likely that irradiation-driven mass-loss has
played a significant role in the evolution of \hatcurb\,--we find that
the planet may have lost $\sim 30\%$ of its present-day mass over the
course of its history, though this conclusion depends strongly on a
number of very poorly constrained parameters, particularly the XUV
flux of \hatcur\ and its evolution in time. We place a 95\% confidence
upper limit on the eccentricity of $e < 0.22$. If further observations
detect a nonzero eccentricity, it would mean that at least three of
the four known Neptune-mass TEPs have nonzero eccentricities, which
may imply that the tidal quality factor is higher than expected for
these planets. Observations of the planetary occultation event for
\hatcurb\ with IRAC/{\em Spitzer} would greatly constrain the
eccentricity, however the low expected depth is likely to make this a
challenging observation. We find suggestive evidence for a linear
drift in the RVs which is significant at the $2.1\sigma$ level. If
confirmed, this would imply the existence of at least one additional
body in the \hatcur\ system. With an expected R-M amplitude of $\sim
9$\,\ms\ and a low stellar RV jitter, \hatcurb\ is a good target to
measure the R-M effect and thereby test the hypothesis that low-mass
planets are more likely to be misaligned than high-mass planets. The
low surface gravity also makes \hatcurb\ a good target for
transmission spectroscopy.

\acknowledgements 

HATNet operations have been funded by NASA grants NNG04GN74G,
NNX08AF23G and SAO IR\&D grants.  Work of G.\'A.B.~and J.~Johnson were
supported by the Postdoctoral Fellowship of the NSF Astronomy and
Astrophysics Program (AST-0702843 and AST-0702821, respectively).  GT
acknowledges partial support from NASA grant NNX09AF59G.  We
acknowledge partial support also from the Kepler Mission under NASA
Cooperative Agreement NCC2-1390 (D.W.L., PI).  G.K.~thanks the
Hungarian Scientific Research Foundation (OTKA) for support through
grant K-81373.  This research has made use of Keck telescope time
granted through NASA (N018Hr and N167Hr).




\begin{thebibliography}{}

\bibitem[Adams \& Laughlin(2006)]{adams:2006}
Adams, F.~C., \& Laughlin, G.\ 2006, \apj, 649, 1004

\bibitem[Ayers(2009)]{ayers:2009}
Ayers, T.~R.\ 2009, \apj, 696, 1931

\bibitem[Bakos et al.(2004)]{bakos:2004}
 Bakos, G.~\'A., Noyes, R.~W., Kov\'acs, G., Stanek, K.~Z.,
 Sasselov, D.~D., \& Domsa, I.~2004, \pasp, 116, 266

\bibitem[Bakos et al.(2007)]{bakos:2007a}
 Bakos, G.~\'A., et al.~2007, \apj, 670, 826

\bibitem[Bakos et al.(2010)]{bakos:2010} Bakos, G.~{\'A}., et al.~2010,
\apj, 710, 1724

\bibitem[Baraffe et al.(1998)]{baraffe:1998} Baraffe, I., Chabrier, G., 
Allard, F., \& Hauschildt, P.~H.~1998, \aap, 337, 403 

\bibitem[Baraffe(2005)]{baraffe:2005}
Baraffe, I., Chabrier, G., Barman, T.~S., Selsis, F., Allard, F., \& Hauschildt, P.~H.\ 2005, \aap, 436, L47

\bibitem[Baraffe et al.(2008)]{baraffe:2008}
Baraffe, I., Chabrier, G., \& Barman, T.~2008, \aap, 482, 315 

\bibitem[Borucki et al.(2010)]{borucki:2010} Borucki, W.~J., et al.\  2010, \apj, 713, L126

\bibitem[Brown(2001)]{brown:2001}
Brown, T.~M.\ 2001, \apj, 553, 1006

\bibitem[Buchhave et al.(2010)]{buchhave:2010}
Buchhave, L.~A., et al.\ 2010, \apj, 720, 1118

\bibitem[Butler et al.(1996)]{butler:1996} 
Butler, R.~P.~et al.~1996, \pasp, 108, 500

\bibitem[Butler et al.(2004)]{butler:2004} Butler, R.~P., Vogt, S.~S.,
  Marcy, G.~W., Fischer, D.~A., Wright, J.~T., Henry, G.~W., Laughlin,
  G., \& Lissauer, J.~J.~2004, \apj, 617, 580

\bibitem[Carpenter(2001)]{carpenter:2001} Carpenter, J.~M.~2001, \aj, 121, 2851 

\bibitem[Charbonneau et al.(2002)]{charbonneau:2002} Charbonneau, D., 
Brown, T.~M., Noyes, R.~W., \& Gilliland, R.~L.~2002, \apj, 568, 377 

\bibitem[Charbonneau et al.(2009)]{dc:2009} Charbonneau, D., et  al.\ 2009, \nat, 462, 891

\bibitem[Claret(2004)]{claret:2004}
 Claret, A.~2004, \aap, 428, 1001

\bibitem[Demory et  al.(2007)]{demory:2007} Demory, B.-O., et al.\ 2007, \aap, 475, 1125

\bibitem[de Pater \& Lissauer(2001)]{depater:2001} de Pater, I., \& Lissauer, J.~J.\ 2001, Planetary Sciences,  p.~80.~ISBN 0521482194.~Cambridge, UK: Cambridge University Press, December 2001.  

\bibitem[Droege et al.(2006)]{droege:2006}
Droege, T.~F., Richmond, M.~W., \& Sallman, M.~2006, \pasp, 118, 1666

\bibitem[Erkaev et al.(2007)]{erkaev:2007} Erkaev, N.~V., Kulikov,
  Y.~N., Lammer, H., Selsis, F., Langmayr, D., Jaritz, G.~F., \&
  Biernat, H.~K.\ 2007, \aap, 472, 329

\bibitem[Etzel(1981)]{etzel:1981}
Etzel, P.~B.\ 1981, NATO ASI, p. 111

\bibitem[Ford(2006)]{ford:2006}
Ford, E.~2006, \apj, 642, 505

\bibitem[Fortney et al.(2007)]{fortney:2007} Fortney, J.~J., Marley, M.~S., \& Barnes, J.~W.~2007, \apj, 659, 1661

\bibitem[F\H ur\'esz(2008)]{furesz:2008} F\H ur\'esz, G.\ 2008, Ph.D. thesis, University of Szeged, Hungary

\bibitem[Gillon et al.(2007)]{gillon:2007} Gillon, M., et al.~2007, \aap, 472, L13

\bibitem[Girardi et al.(2000)]{girardi:2000} Girardi, L., Bressan, A., 
Bertelli, G., \& Chiosi, C.~2000, \aaps, 141, 371 

\bibitem[Hansen \& Barman(2007)]{hansen:2007} Hansen, B.~M.~S., \& Barman, T.~2007, \apj, 671, 861 

\bibitem[Hartman et al.(2009)]{hartman:2009} Hartman, J.~D., et al.\ 2009, \apj, 706, 785

\bibitem[Hartman et al.(2010)]{hartman:2010} Hartman, J.~D., et al.\ 2010, \apj, submitted, arXiv:1007.4850

\bibitem[Hirano et al.(2010)]{hirano:2010} Hirano, T., Narita, N.,
  Shporer, A., Sato, B., Aoki, W., \& Tamura, M.\ 2010, \pasj,
  submitted, arXiv:1009.5677

\bibitem[Holman et al.(2010)]{holman:2010} Holman, M.~J., et al.\ 2010, Science, in press

\bibitem[Isaacson \& Fischer(2010)]{isaacson:2010}
Isaacson, H., \& Fischer, D.~A.\ 2010, \apj, in press, arXiv:1009.2301

\bibitem[Jackson et al.(2010)]{jackson:2010} Jackson, B., Miller, N.,
  Barnes, R., Raymond, S.~N., Fortney, J.~J., \& Greenberg, R.\ 2010,
  \mnras, 407, 910

\bibitem[Kipping  \& Bakos(2010a)]{kipping:2010} Kipping, D.~M., \& Bakos, G.~{\'A}.\ 2010a, arXiv:1004.3538

\bibitem[Kipping \& Bakos(2010b)]{kipping:2010b} Kipping, D.~M., \& Bakos, G.~{\'A}.\ 2010b, arXiv:1006.5680

\bibitem[Kipping et al.(2010)]{kipping:2010c} Kipping, D.~M., et al.\ 2010, \apj, submitted, arXiv:1008.3389

\bibitem[Knutson et al.(2009)]{knutson:2009}
Knutson, H.~A., Charbonneau, D., Burrows, A., O'Donovan, F.~T., \& Mandushev, G.\ 2009, \apj, 691, 866

\bibitem[Komatsu et al.(2010)]{komatsu:2010}
Komatsu, E., et al.\ 2010, \apjs\ submitted, arXiv:1001.4538

\bibitem[Kov\'acs et al.(2002)]{kovacs:2002}
Kov\'acs, G., Zucker, S., \& Mazeh, T.~2002, \aap, 391, 369

\bibitem[Kov\'acs et al.(2005)]{kovacs:2005}
Kov\'acs, G., Bakos, G.~\'A., \& Noyes, R.~W.~2005, \mnras, 356, 557

\bibitem[Kurtz(1985)]{kurtz:1985}
Kurtz, D.~W.\ 1985, \mnras, 213, 773

\bibitem[Lammer et al.(2003)]{lammer:2003}
Lammer, H., Selsis, F., Ribas, I., Guinan, E.~F., Bauer, S.~J., \& Weiss, W.~W.\ 2003, \apjl, 598, L121

\bibitem[Lecavelier des Etangs et al.(2010)]{lecavelierdesetangs:2010}
Lecavelier des Etangs, A., et al.\ 2010, \aap, in press, arXiv:1003.2206

\bibitem[L\'eger et  al.(2009)]{leger:2009} L{\'e}ger, A., et al.\ 2009, \aap, 506, 287

\bibitem[Lucy \& Sweeney(1971)]{lucy:1971}
Lucy, L.~B., \& Sweeney, M.~A.\ 1971, \aj, 76, 544

\bibitem[Lupton(1993)]{lupton:1993} Lupton, R.\ 1993, ``Statistics in
  Theory and Practice'', Princeton N.J.: Princeton University Press,
  p. 100

\bibitem[Mamajek \& Hillenbrand(2008)]{mamajek:2008}
Mamajek, E.~E., \& Hillenbrand, L.~A.\ 2008, \apj, 687, 1264

\bibitem[Mandel \& Agol(2002)]{mandel:2002}
 Mandel, K., \& Agol, E.~2002, \apjl, 580, L171

\bibitem[Marcy \& Butler(1992)]{marcy:1992}
 Marcy, G.~W., \& Butler, R.~P.~1992, \pasp, 104, 270

\bibitem[Nelson \& Davis(1972)]{nelson:1972}
Nelson, B., \& Davis, W.~D.\ 1972, \apj, 174, 617

\bibitem[Noyes et al.(1984)]{noyes:1984} Noyes, R.~W., Hartmann, L.~W.,
Baliunas, S.~L., Duncan, D.~K., \& Vaughan, A.~H.~1984, \apj, 279, 763

\bibitem[P\'al \& Bakos(2006)]{pal:2006}
 P\'al, A., \& Bakos, G.~\'A. 2006, \pasp, 118, 1474

\bibitem[P{\'a}l et al.(2008)]{pal:2008} P{\'a}l, A., et al.~2008, 
\apj, 680, 1450

\bibitem[P\'al(2009b)]{pal:2009b}
P\'al, A.\ 2009b, arXiv:0906.3486, PhD thesis

\bibitem[Popper \& Etzel(1981)]{popper:1981}
Popper, D.~M., \& Etzel, P.~B.\ 1981, \aj, 86, 102

\bibitem[Press et al.(1992)]{press:1992}
Press, W. H., Teukolsky, S. A., Vetterling, W. T. \& Flannery, B. P., 1992,
Numerical  Recipes in C: the art of scientific computing,
Second Edition, Cambridge University Press

\bibitem[Queloz et al.(2001)]{queloz:2001}
Queloz, D.~et al.~2001, \aap, 379, 279

\bibitem[Queloz et  al.(2009)]{queloz:2009} Queloz, D., et al.\ 2009, \aap, 506, 303

\bibitem[Quinn et al.(2010)]{quinn:2010}
Quinn, S.~N., et al.\ 2010, \apj submitted, arXiv:1008.3565

\bibitem[Ram{\'{\i}}rez \& Mel{\'e}ndez(2005)]{ramirez:2005}
Ram{\'{\i}}rez, I., \& Mel{\'e}ndez, J.~2005, \apj, 626, 465

\bibitem[Ribas et al.(2005)]{ribas:2005} Ribas, I., Guinan, E.~F., G\"udel, M., \& Audard, M.\ 2005, \apj, 622, 680

\bibitem[Roeser et al.(2010)]{roeser:2010} Roeser, S., Demleitner, M., \& Schilbach, E.\ 2010, \aj, 139, 2440

\bibitem[Schlaufman(2010)]{schlaufman:2010} Schlaufman, K.~C.\ 2010, \apj, 719, 602

\bibitem[Seager \& Mall\'en-Ornelas(2003)]{seager:2003}
Seager, S., \& Mall\'en-Ornelas, G.\ 2003, \apj, 585, 1038

\bibitem[Seager \& Deming(2010)]{seager:2010}
Seager, S., \& Deming, D.\ 2010, \araa, 48, 631

\bibitem[Seidelmann et al.(2007)]{seidelmann:2007}
Seidelmann, P.~K., et al.\ 2007, Celest. Mech. Dyn. Astron., 98, 155

\bibitem[Sing et al.(2010)]{sing:2010} Sing, D.~K.\ 2010, \aap\ submitted, arXiv:1008.4795
 
\bibitem[Skrutskie et al.(2006)]{skrutskie:2006} Skrutskie, M.~F., et 
al.~2006, \aj, 131, 1163

\bibitem[Southworth et al.(2004a)]{southworth:2004a} Southworth, J.,
  Maxted, P.~F.~L., \& Smalley, B.\ 2004a, \mnras, 351, 1277

\bibitem[Southworth et al.(2004b)]{southworth:2004b} Southworth, J.,
  Zucker, S., Maxted, P.~F.~L., \& Smalley, B.\ 2004b, \mnras, 355,
  986

\bibitem[Southworth(2009)]{southworth:2009} Southworth, J.~2008, \mnras, 394, 272

\bibitem[Sozzetti et al.(2007)]{sozzetti:2007}
 Sozzetti, A.~et al.~2007, \apj, 664, 1190

\bibitem[Torres et al.(2005)]{torres:2005}
 Torres, G., Konacki, M., Sasselov, D.~D., \& Jha, S.~2005, \apj, 619, 558

\bibitem[Torres et al.(2007)]{torres:2007}
 Torres, G.~et al.~2007, \apjl, 666, 121

\bibitem[Torres et al.(2010)]{torres:2010}
 Torres, G., et al.\ 2010, \apj, submitted, arXiv:1008.4393

\bibitem[Triaud et al.(2010)]{triaud:2010}
Triaud, A.~H.~M.~J., et al.\ 2010, \aap, in press, arXiv:1008.2353

\bibitem[Valencia et al.(2010)]{valencia:2010}
Valencia, D., Ikoma, M., Guillot, T., \& Nettelmann, N.\ 2010, \aap, 516, A20

\bibitem[Valenti \& Fischer(2005)]{valenti:2005}
 Valenti, J.~A., \& Fischer, D.~A. 2005, \apjs, 159, 141

\bibitem[Valenti \& Piskunov(1996)]{valenti:1996}
 Valenti, J.~A., \& Piskunov, N.~1996, \aaps, 118, 595

\bibitem[Vaughan, Preston \& Wilson(1978)]{vaughan:1978}
Vaughan, A.~H., Preston, G.~W., \& Wilson, O.~C.~1978, \pasp, 90, 267

\bibitem[Vidal-Madjar et al.(2003)]{vidalmadjar:2003}
Vidal-Madjar, A., Lecavelier des Etangs, A., D\'esert, J.-M., Ballester, G.~E., Ferlet, R., H\'ebrard, G., \& Mayor, M.\ 2003, \nat, 422, 143

\bibitem[Vogt et al.(1994)]{vogt:1994}
 Vogt, S.~S.~et al.~1994, Proc.~SPIE, 2198, 362

\bibitem[Winn(2010)]{winn:2010c}
Winn, J.~N.\ 2010, arXiv:1001.2010

\bibitem[Winn et al.(2010a)]{winn:2010a} 
Winn, J.~N., Fabrycky, D., Albrecht, S., \& Johnson, J.~A.\ 2010a, \apj, 718, L145

\bibitem[Winn et al.(2010b)]{winn:2010b}
Winn, J.~N., et al.\ 2010, \apjl, in press, arXiv:1009.5671

\bibitem[Yelle et al.(2008)]{yelle:2008} 
Yelle, R., Lammer, H., \& Ip, W.\ 2008, Space Sci. Rev., 139, 437

\bibitem[Yi et al.(2001)]{yi:2001}
 Yi, S.~K.~et al.~2001, \apjs, 136, 417

\end{thebibliography}
\end{document}